\documentclass[fleqn,usenatbib]{mnras}
\usepackage{newtxtext,newtxmath}
\usepackage[T1]{fontenc}
\usepackage{graphicx}	% Including figure files
\usepackage{amsmath}	% Advanced maths commands
\usepackage{amssymb}	% Extra maths symbols
\usepackage{xspace}	
\usepackage{lineno} %linenumber
\newcommand{\code}[1]{\textsc{#1}\xspace} %MNRAS style is small caps
\newcommand{\gaia}{{\it Gaia}\xspace}
\newcommand{\kms}{\unit{km\,s^{-1}}}
\newcommand{\masyr}{\unit{mas\,yr^{-1}}}

\newcommand{\SSSSS}{${S}^5$\xspace}

\newcommand{\response}[1]{{#1}}

\newcommand{\muone}{\ensuremath{\mu_{\phi_1}}\xspace}
\newcommand{\muonenot}{\ensuremath{\mu_{\phi_{1,0}}\xspace}}
\newcommand{\mutwo}{\ensuremath{\mu_{\phi_2}}\xspace}
\newcommand{\mutwonot}{\ensuremath{\mu_{\phi_{2,0}}\xspace}}
\newcommand{\mui}{\ensuremath{\mu_{\phi_i}}\xspace}
\newcommand{\muinot}{\ensuremath{\mu_{\phi_{i,0}}\xspace}}

\newcommand{\unit}[1]{\ensuremath{\mathrm{\,#1}}\xspace}
\newcommand{\feh}{\unit{[Fe/H]}}
\newcommand{\teff}{\ensuremath{T_\mathrm{eff}}\xspace}
\newcommand{\logg}{\ensuremath{\log\,g}\xspace}
\newcommand{\alphafe}{\unit{[\alpha/Fe]}}

\title[\SSSSS: Survey Overview]{The Southern Stellar Stream Spectroscopic Survey (\SSSSS): \\ Overview, Target Selection, Data Reduction, Validation, and Early Science}

\author[Li et al.]{
\parbox{\textwidth}{
\Large
T.~S.~Li$^{1,2,3,4}$\thanks{Email: tingli@carnegiescience.edu; NHFP Einstein Fellow},
S.~E.~Koposov$^{5,6}$,
D.~B.~Zucker$^{7,8}$,
G.~F.~Lewis$^{9}$,
K.~Kuehn$^{10,11}$,
J.~D.~Simpson$^{12}$,
A.~P.~Ji$^{1}$\thanks{Hubble Fellow},
N.~Shipp$^{13,4,3}$,
Y.-Y.~Mao$^{14,15,16}$\thanks{NHFP Einstein Fellow},
M.~Geha$^{17}$,
A.~B.~Pace$^{18}$,
A.~D.~Mackey$^{19}$,
S.~Allam$^{3}$,
D.~L.~Tucker$^{3}$,
G.~S.~Da~Costa$^{19}$,
D.~Erkal$^{20}$,
J.~D.~Simon$^{1}$,
J.~R.~Mould$^{21}$,
S.~L.~Martell$^{12,22}$,
Z.~Wan$^{9}$,
G.~M.~De~Silva$^{11}$,
K.~Bechtol$^{23}$,
E.~Balbinot$^{24}$,
V.~Belokurov$^{6}$,
J.~Bland-Hawthorn$^{9,22}$,
A.~R.~Casey$^{25}$,
L.~Cullinane$^{19}$,
A.~Drlica-Wagner$^{3,13,4}$,
S.~Sharma$^{9,22}$,
A.~K.~Vivas$^{26}$,
R.~H.~Wechsler$^{27,28,29}$,
B.~Yanny$^{3}$
\begin{center} (\SSSSS Collaboration) \end{center}
{\it(Affiliations are listed after the references)}\\
}
}

\date{Accepted XXX. Received YYY; in original form ZZZ}

\pubyear{2018}

\begin{document}

%\linenumbers

%%%%%%%%%%%%Fermilab Publication Number%%%%%%%%%
%%
%%%FERMILAB-PUB-19-290-AE
%%
%%%%%%%%%%%%%%%%%%%%%%%%%%%%%%%%%%%%%%%%

\label{firstpage}
\pagerange{\pageref{firstpage}--\pageref{lastpage}}
\maketitle

\begin{abstract}
We introduce the Southern Stellar Stream Spectroscopy Survey (\SSSSS), an on-going program to map the kinematics and chemistry of stellar streams in the Southern Hemisphere. The initial focus of \SSSSS has been spectroscopic observations of recently identified streams within the footprint of the Dark Energy Survey (DES), with the eventual goal of surveying streams across the entire southern sky.
Stellar streams are composed of material that has been tidally striped from dwarf galaxies and globular clusters and hence are excellent dynamical probes of the gravitational potential of the Milky Way, as well as providing a detailed snapshot of its accretion history. Observing with the 3.9-m Anglo-Australian Telescope's 2-degree-Field fibre positioner and AAOmega spectrograph, and combining the precise photometry of DES DR1 with the superb proper motions from \gaia DR2, allows us to conduct an efficient spectroscopic survey to map these stellar streams. So far \SSSSS has mapped 9 DES streams and 3 streams outside of DES; the former are the first spectroscopic observations of these recently discovered streams. In addition to the stream survey, we use spare fibres to undertake a Milky Way halo survey and a low-redshift galaxy survey. This paper presents an overview of the \SSSSS program, describing the scientific motivation for the survey, target selection, observation strategy, data reduction and survey validation.  
Finally, we describe early science results on stellar streams and Milky Way halo stars drawn from the survey. \response{Updates on \SSSSS, including future public data releases, can be found at \url{http://s5collab.github.io}}.
\end{abstract}


%%%%%%%%%%%%%%%%% BODY OF PAPER %%%%%%%%%%%%%%%%%%

\section{Introduction}\label{sec:intro}

Within the $\Lambda$CDM cosmological model, large galaxies grow hierarchically through the accretion of smaller systems. In the inner parts of galaxies, where dynamical time-scales are relatively short, these accreted systems are rapidly phase-mixed into a comparatively smooth stellar halo. However, in the outer stellar halo, where dynamical time-scales are longer, accreted systems are only partially phase mixed, exhibiting the signatures of ongoing tidal disruption. Hence, the distribution of stellar debris in the halo provides a snapshot of the galactic evolution of our Milky Way \citep[][]{2002ARA&A..40..487F,2005ApJ...635..931B}.

The structural and kinematic properties of tidal stellar streams also provide a measurement of the mass and shape of the Milky Way's dark matter halo. While this dark matter dominates the gravitational potential of the Milky Way, there remain significant uncertainties in its properties, limiting the accuracy of comparisons to predictions of hierarchical structure formation models. Hence, modelling the dynamical properties of a large sample of stellar streams, spread over a broad range of Galactocentric distances, offers the realistic prospects of accurately determining Galaxy's gravitational potential \citep[e.g.][]{Johnston:1999,Bonaca:2018}.

Over recent years, there have been significant efforts to uncover stellar substructure in our Galactic halo, with more than fifty stellar streams now known, half of which were discovered in the last three years \citep[][and references therein]{Mateu:2017}. 
In particular, the Dark Energy Survey (DES), with its unprecedented photometric calibration, depth, and sky coverage, 
has recently recovered four previously known stellar streams \citep{Koposov2014, Drlica-Wagner:2015, Balbinot:2016, Grillmair:2017} and discovered eleven new streams in the Southern sky through isochrone matching of metal-poor populations throughout the stellar halo \citep{Shipp:2018}.
While imaging surveys, such as DES, can provide on-sky locations and distance estimates through isochrone fitting, spectroscopy is essential for measuring the kinematic and chemical properties of stream stars, allowing the determination of radial velocities, velocity dispersions, and gradients; this information is required to deduce the dynamical history of a stellar stream and infer the three-dimensional structure of the Milky Way's dark matter halo \citep[e.g.][]{1998ApJ...500..575I,2001ApJ...551..294I,Koposov:2010,2011MNRAS.417..198V,Gibbons:2014,Bowden:2015,Erkal:2016,Bovy:2016,Bonaca:2018,Erkal:2019b}. Spectroscopy is also crucial when using streams to measure the properties of dark matter subhaloes \citep[e.g.][]{Ibata:2002,Johnston:2002} since subhalo impacts create correlated signals in all of the stream observables \citep[e.g.][]{Yoon:2011,Carlberg:2013,Erkal:2015a,Helmi:2016,Sanders:2016,Bovy:2017} and at least three observables are needed to recover the subhalo properties \citep{Erkal:2015}. 

Spectroscopy of stellar streams is challenging due to the relative faintness of stream-member stars ($g \sim 19$, for a horizontal branch star at 45 kpc), the low stellar surface density, with only several stars per deg$^2$ at $g \sim 19$, and substantial contamination from Milky Way foreground stars, with hundreds per deg$^2$ at $g \sim 19$. 
Despite the rapid increase in the number of known streams, these observational challenges have limited their detailed spectroscopic investigation, and hence their use as cosmological probes \citep[see e.g.,][]{Majewski:2004, Koposov:2010, Sesar:2015, Ibata:2016}. In order to investigate accretion processes and progenitors, we place a premium on assembling a large sample of streams, a large sample of stars per stream, and accurate kinematics, in the expectation that stream kinematics, including
internal kinematics, retain a memory of initial conditions.

The Southern Stellar Stream Spectroscopic Survey (\SSSSS) was initiated in mid-2018 to address the challenges associated with spectroscopic observations of stellar streams. 
To date, \SSSSS represents the first spectroscopic survey of stellar streams in our Galactic halo. 
\SSSSS uses the Two-degree Field (2dF) fibre positioner \citep{Lewis:2002} coupled with the dual-arm AAOmega spectrograph \citep{Sharp:2006} on the 3.9-m Anglo-Australian Telescope (AAT); 2dF provides 392 science fibres that can be distributed across a field of view (FOV) of $\sim3$ deg$^2$. \SSSSS is an ongoing survey, with 25 nights observed in 2018 and 12 hrs observed in 2019 as of June 2019, and more nights planned in 2019. 
Though \SSSSS intends to expand the targeted streams to the entire Southern Sky, our 2018 observations primarily targeted streams in the DES footprint. Therefore, this paper will mainly focus on the target selection and observations of the 14 DES streams.\footnote{There are 15 streams identified in the DES footprint~\citep{Shipp:2018}. The Palca stream is not considered in \SSSSS due to its low surface brightness and diffuse morphology.}

The target selection for \SSSSS uses the recently released parallax and proper motion information from \gaia DR2 \citep{GaiaCollaboration:2016cu,Gaia:2018}, together with precise photometry from the latest data releases of ground-based imaging surveys, mainly DES DR1 \citep{desdr1}. Although 2dF provides substantial spectroscopic multiplexing, the diffuse nature of stellar streams still requires efficient target selection.
Fortunately, proper motions from \gaia DR2 have dramatically improved the target selection efficiency of stream candidates, which allows us to conduct two auxiliary science programs with spare fibres: a Milky Way halo star survey and a low redshift (low-z) galaxy survey. 
While \SSSSS is mainly focused on stellar streams, this paper also provides an overview of the experimental design, target selection and data reduction of, and some early science from those auxiliary surveys.

The structure of this paper is as follows: 
Section~\ref{sec:selection} presents the details of the field and target selection for \SSSSS, while Section~\ref{sec:observations} details the observational program and subsequent data reduction. 
Section~\ref{sec:validation} lays out the validation of the survey, followed by a discussion of early science results in  Section~\ref{sec:science}. Our conclusions and plans for the future of S5 are presented in Section~\ref{sec:summary}.

We note that in this paper, we use lower case $griz$ for DECam photometry (except for Section \ref{sec:emp} on SkyMapper photometry), where the photometry comes from either DES DR1 or the Dark Energy Camera Legacy Survey (DECaLS) DR7 \citep{Dey:2018}, and we use $G, G_\textrm{BP}, G_\textrm{RP}$ for \gaia photometry. We use the subscript 0 to denote the reddening-corrected photometry throughout the paper. For DECam photometry, the reddening correction was performed following the procedures described in DES DR1. Specifically, we calculated the extinction by multiplying the colour excess $E(B-V)$ from \citet{Schlegel:1998} with the extinction coefficients taken from DES DR1 \citep{desdr1}. For the \gaia photometry, we use the colour-dependent extinction corrections from \citet{Babusiaux2018} and the \citet{Schlegel:1998} values of $E(B-V)$.

\section{Survey Design and Target Selection}\label{sec:selection}

In this section, we first define the AAT fields for \SSSSS (Section \ref{sec:fields}), and then we discuss the target selection for each field, including stream targets, halo targets and the low-z targets. The 2dF fibre allocation software \code{configure}\footnote{https://www.aao.gov.au/science/software/configure} \citep{Miszalski:2006ef} allows the targets to be given a priority in the range 1--9 (P1--P9, with P9 being the highest).  The higher the target priority, the more likely it is to be allocated a fibre by \code{configure}.  We therefore assigned our stream targets to the highest priority range (P9--P7), halo targets to the next priority range (P6--P3), and the low-z galaxy targets to the lowest priorities (P2--P1). 
In Table \ref{table:targets}, we summarise the targets for each priority category. We detail the stream targets in Section \ref{sec:stream_target}, non-stream stellar targets in Section \ref{sec:other_targets}, and galaxy targets in Section \ref{sec:galaxy_targets}.
Some targets could fall in multiple priority categories; in those cases the highest priority was assigned.

\begin{table*}
\begin{center}
\caption{
Description of P9 to P1 targets for \SSSSS.
}
\label{table:targets}
\begin{tabular}{c l}
\hline
Priority  & Target Description \\
\hline
\hline
P9   &   Stream candidates: metal-poor, PM1 (tight PM cut)   \\
P8   &   Stream candidates: PM2 (less tight PM cut)   \\
P7   &   Stream candidates: metal-poor, PM3 (loose PM cut)   \\
\hline
P6   &   Rare objects (BHBs, RRLs, WDs)  \\
P5   &   Extreme metal-poor star candidates (SkyMapper photometry)  \\
P4   &   Metal-poor stars (DES photometry) \\
P3   &   Low PM stars \\
\hline
P2   &   High-probability low-z galaxy candidates \\
P1   &   Low-probability low-z galaxy candidates \\
\hline
\end{tabular}
\end{center}
\end{table*}

\subsection{Field Selection}\label{sec:fields}
In defining the AAT fields for \SSSSS, and considering the FOV of 2dF, the goal of our survey design is to cover the maximum sky area within the limited amount of total telescope time available. 
In summary, we separate AAT pointings by 2\degr, just under the diameter of the 2dF FOV, aligned with each stream's ridgeline. For most streams, the ridgeline is defined as the heliocentric great circle from the end points of the stream defined in~\citet{Shipp:2018}. For these streams, we make the field grid in the stream coordinates along stream latitude $\phi_2 = 0\degr$ and stream longitude $\phi_1 = ..., -2\degr, 0\degr, 2\degr...$, and then we transfer from stream coordinates to celestial coordinates using the rotation matrix for each stream.\footnote{The rotation matrices are defined in the Appendix C of Shipp et al. (\textit{submitted\@}).} The only exception is the ATLAS stream, for which~\citet{Shipp:2018} found that the stream ridgeline deviates significantly from a great circle. We therefore used the polynomial track defined in~\citet{Shipp:2018} as the ridgeline for the ATLAS stream in our field definition. 

The number of fields used for each stream depends on the length of the stream. For most streams, we have $L/2$ (rounded to the nearest integer) AAT fields, where $L$ is the length of the stream in degrees from \citet{Shipp:2018}. For some streams, we obtained 1--2 extra AAT fields extending from the endpoints of the streams, to search for possible members beyond the photometric extent. 

\begin{figure*}
\centerline{\includegraphics[width=5.5in]{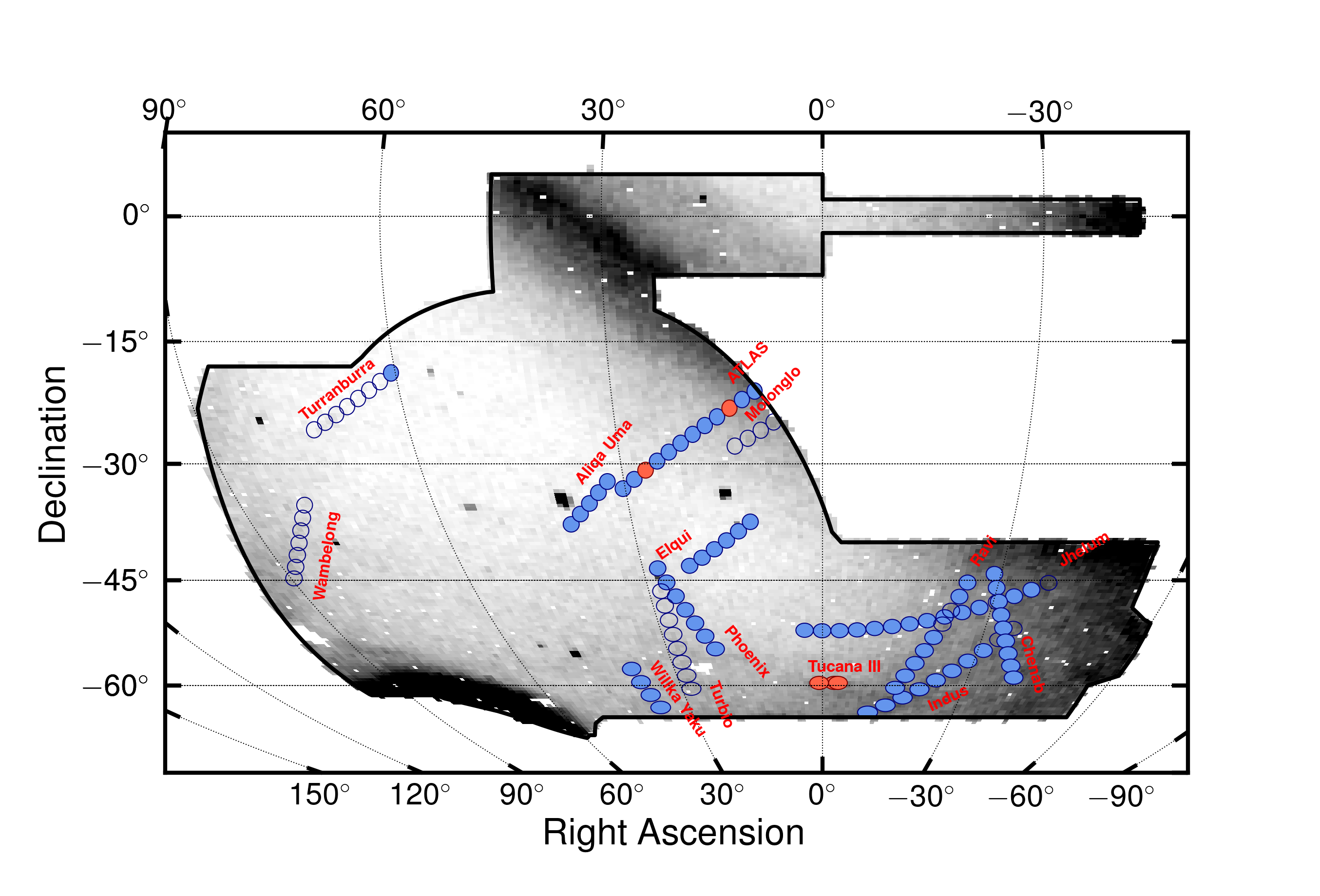}}
\caption{
Pointing and status map of \SSSSS within the footprint of DES. Each circle shows one AAT pointing, with filled blue ones for observations accomplished so far (as of June 2019), filled red ones for observations taken prior to \SSSSS, and open circles are the remaining fields to be observed in 2019.
The background 2D histogram shows the stellar count density of main-sequence stars at distance modulus $m-M$=16.8 in the DES footprint.
}
\label{fig:s5pointing}
\end{figure*}

An illustration of the stream fields is shown in Figure~\ref{fig:s5pointing} and the centres of the fields are listed in Table~\ref{table:fields}. Among the 14 DES streams, 10 streams have more than 80\% of their observations completed to date: Tucana III, ATLAS, Aliqa Uma, Chenab, Elqui, Jhelum, Indus, Phoenix, Ravi, and Willka Yaku. The other four streams are planned for observation in 2019.

Before the start of \SSSSS, we carried out  pilot programs on some of the stream fields, shown as the red filled circles in Figure~\ref{fig:s5pointing}. The Tucana III stream (at $\alpha_{2000} \sim 0\degr$ and $\delta_{2000} \sim -60\degr$) was observed in 2016 and was published in \citet{Li:2018b}. Two fields in the ATLAS stream were observed in 2018A. Proper motions from \gaia were not available then and therefore the target selection strategy described below does not apply to those pre-\SSSSS fields. However, data collected from these two ATLAS fields are still considered as part of \SSSSS in the data reduction and final catalogue production since the instrument settings were the same. 

In addition, a few streams outside of the DES footprint
were observed. The field selection presented here, as well as the target selection strategy described below, do not apply to those non-DES fields and we discuss them in Section \ref{sec:nonDES}.

\begin{table*}
\begin{center}
\caption{
\SSSSS fields observed with the AAT as of June 2019. Columns from left to right are field name, RA and Decl. of the centre of the fields, UT date of the observation (with highest S/N if observed multiple times), MJD (start) of the observation, total exposure time (in seconds), average \gaia $G-$band magnitude at S/N = 5 per pixel \response{(from red arm spectra)}, total number of targets ($N_\mathrm{targets}$), \response{and number of stars ($N_\mathrm{goodstar}$) with good measurements (i.e. \code{good\_star} = 1; see definition in Section \ref{sec:goodstar}). We note that for fields in some streams such as ATLAS, Elqui, Phoenix, etc., $N_\mathrm{goodstar}$ is usually much lower than $N_\mathrm{targets}$, because these streams are at high Galactic latitude; therefore, we used $\sim100$ spare fibres to target low redshift galaxies (see Section~\ref{sec:galaxy_targets}), and all galaxy targets are assigned \code{good\_star} = 0 regardless of the quality of the spectra.} 
All fields are grouped into four categories, which are fields in the DES footprint, fields outside the DES footprint (see Section \ref{sec:nonDES}), calibration fields for survey validation (see Section \ref{sec:validation}), and fields observed prior to \SSSSS with the same instrument setup but previously unpublished. 
}
\scriptsize
\label{table:fields}
\begin{tabular}{l r r l l c c r r}
\hline
Field Name  & RA (deg)  & Decl. (deg) & UTDATE & MJD & $t_\mathrm{exp}$ (s) & $G$@(S/N=5) & $N_\mathrm{targets}$ & $N_\mathrm{goodstar}$ \\
\hline
\hline

             ATLAS-0 &  30.350248  & -33.098693  & 2018-09-14   & 58375.59  &  7200  & 18.9  & 347  & 182 \\
             ATLAS-1 &  28.406544  & -31.922932  & 2018-09-13   & 58374.60  &  7200  & 19.1  & 359  & 188 \\
             ATLAS-3 &  24.569248  & -29.628485  & 2018-09-12   & 58373.62  &  7200  & 18.8  & 359  & 187 \\
             ATLAS-4 &  22.671638  & -28.507177  & 2018-09-27   & 58388.69  &  7200  & 18.6  & 359  & 140 \\
             ATLAS-5 &  20.784285  & -27.396658  & 2018-10-26   & 58417.68  &  5800  & 18.4  & 345  & 153 \\
             ATLAS-6 &  18.912949  & -26.299787  & 2018-09-11   & 58372.59  &  7200  & 18.7  & 359  & 185 \\
             ATLAS-7 &  17.046881  & -25.215898  & 2018-10-23   & 58414.54  &  7200  & 18.4  & 359  & 162 \\
             ATLAS-8 &  15.186390  & -24.141606  & 2018-10-26   & 58417.59  &  7200  & 18.5  & 359  & 156 \\
            ATLAS-10 &  11.485245  & -22.025185  & 2018-09-08   & 58369.59  &  7200  & 18.7  & 359  & 185 \\
            ATLAS-11 &   9.638663  & -20.985328  & 2018-09-27   & 58388.60  &  7200  & 18.5  & 348  & 165 \\
         Aliqa-Uma-1 &  39.653899  & -37.658852  & 2018-09-08   & 58369.69  &  7800  & 18.7  & 347  & 222 \\
         Aliqa-Uma-2 &  37.776994  & -36.334496  & 2018-09-07   & 58368.70  &  7600  & 18.8  & 347  & 215 \\
         Aliqa-Uma-3 &  35.965770  & -34.980462  & 2018-09-09   & 58370.67  &  9600  & 19.0  & 347  & 219 \\
         Aliqa-Uma-4 &  34.211411  & -33.603704  & 2018-08-09   & 58339.73  &  7200  & 19.1  & 348  & 211 \\
         Aliqa-Uma-5 &  32.515170  & -32.198125  & 2018-08-08   & 58338.75  &  5400  & 18.8  & 308  & 191 \\
            Chenab-1 & 331.107304  & -44.179689  & 2018-08-10   & 58340.46  &  7200  & 19.0  & 359  & 340 \\
            Chenab-2 & 330.109770  & -46.053830  & 2018-08-10   & 58340.55  &  6700  & 18.6  & 348  & 288 \\
            Chenab-3 & 329.045873  & -47.914236  & 2018-09-07   & 58368.41  &  7200  & 18.5  & 357  & 311 \\
            Chenab-4 & 327.900452  & -49.766183  & 2018-09-11   & 58372.40  &  7200  & 18.9  & 359  & 328 \\
            Chenab-5 & 326.663567  & -51.606374  & 2018-09-09   & 58370.39  &  7200  & 18.7  & 359  & 345 \\
            Chenab-6 & 325.321207  & -53.432514  & 2018-09-08   & 58369.40  &  7200  & 18.6  & 359  & 341 \\
            Chenab-7 & 323.861828  & -55.242360  & 2018-09-13   & 58374.42  &  7200  & 19.1  & 359  & 333 \\
            Chenab-8 & 322.259995  & -57.036550  & 2018-08-09   & 58339.55  &  6300  & 19.0  & 348  & 324 \\
            Chenab-9 & 320.500812  & -58.803070  & 2018-08-09   & 58339.46  &  6000  & 18.8  & 359  & 315 \\
             Elqui-0 &  22.107923  & -43.089436  & 2018-09-14   & 58375.68  &  7800  & 18.7  & 359  & 187 \\
             Elqui-1 &  19.813988  & -42.021790  & 2018-09-10   & 58371.59  &  9000  & 19.0  & 359  & 235 \\
             Elqui-2 &  17.594839  & -40.909998  & 2018-08-09   & 58339.63  &  7800  & 19.0  & 348  & 166 \\
             Elqui-3 &  15.453303  & -39.757473  & 2018-08-08   & 58338.66  &  7200  & 18.9  & 348  & 177 \\
             Elqui-4 &  13.383956  & -38.560740  & 2018-09-09   & 58370.55  &  9000  & 18.8  & 359  & 187 \\
             Elqui-5 &  11.380964  & -37.334282  & 2018-08-07   & 58337.59  &  7500  & 18.9  & 333  & 189 \\
             Indus-1 & 349.678452  & -64.192996  & 2018-08-11   & 58341.66  &  3970  & 17.6  & 347  & 193 \\
             Indus-2 & 345.944962  & -63.071643  & 2018-08-02   & 58332.73  &  7200  & 18.1  & 358  & 240 \\
             Indus-3 & 342.498793  & -61.863658  & 2018-08-02   & 58332.66  &  5400  & 18.5  & 348  & 299 \\
             Indus-4 & 339.324859  & -60.573093  & 2018-08-12   & 58342.61  &  5720  & 18.3  & 344  & 307 \\
             Indus-5 & 336.397623  & -59.212635  & 2018-08-11   & 58341.55  &  7760  & 18.1  & 359  & 297 \\
             Indus-6 & 333.700573  & -57.792776  & 2018-08-04   & 58334.74  &  6300  & 18.3  & 359  & 305 \\
             Indus-7 & 331.210566  & -56.320239  & 2018-08-02   & 58332.59  &  5400  & 18.5  & 359  & 308 \\
             Indus-8 & 328.910308  & -54.804194  & 2018-08-02   & 58332.52  &  5400  & 18.5  & 348  & 323 \\
            Jhelum-1 &   3.341637  & -51.908154  & 2018-09-12   & 58373.53  &  7200  & 18.4  & 347  & 275 \\
            Jhelum-2 &   0.100185  & -51.978971  & 2018-09-09   & 58370.48  &  5400  & 18.5  & 347  & 280 \\
            Jhelum-3 & 356.853568  & -51.960940  & 2018-09-10   & 58371.49  &  7200  & 18.5  & 347  & 285 \\
            Jhelum-4 & 353.614405  & -51.853841  & 2018-10-23   & 58414.44  &  7200  & 18.2  & 349  & 294 \\
            Jhelum-5 & 350.398170  & -51.660569  & 2018-08-04   & 58334.67  &  6000  & 18.6  & 348  & 292 \\
            Jhelum-6 & 347.216054  & -51.376758  & 2018-08-04   & 58334.60  &  5400  & 18.6  & 359  & 298 \\
            Jhelum-7 & 344.078713  & -51.011100  & 2018-08-01   & 58331.64  &  7200  & 18.8  & 348  & 295 \\
            Jhelum-8 & 340.996359  & -50.560674  & 2018-09-13   & 58374.51  &  6600  & 18.4  & 347  & 321 \\
            Jhelum-9 & 337.979644  & -50.031113  & 2018-09-10   & 58371.40  &  7200  & 18.7  & 359  & 351 \\
           Jhelum-10 & 335.027230  & -49.426959  & 2018-09-08   & 58369.50  &  7200  & 18.7  & 347  & 340 \\
           Jhelum-11 & 332.150927  & -48.748199  & 2018-10-25   & 58416.54  &  5400  & 17.8  & 349  & 258 \\
           Jhelum-13 & 326.650541  & -47.188388  & 2018-08-08   & 58338.60  &  4800  & 18.1  & 359  & 340 \\
           Jhelum-14 & 324.025939  & -46.315883  & 2018-08-01   & 58331.51  &  7200  & 18.8  & 367  & 340 \\
           Phoenix-1 &  27.522755  & -43.434219  & 2018-09-10   & 58371.70  &  7800  & 18.7  & 347  & 199 \\
           Phoenix-2 &  26.574314  & -45.316320  & 2018-08-01   & 58331.73  &  7200  & 18.8  & 359  & 179 \\
           Phoenix-3 &  25.563754  & -47.189449  & 2018-09-07   & 58368.60  &  7800  & 18.6  & 359  & 214 \\
           Phoenix-4 &  24.478430  & -49.055840  & 2018-08-06   & 58336.61  &  9600  & 18.7  & 348  & 204 \\
           Phoenix-5 &  23.306432  & -50.906988  & 2018-08-07   & 58337.76  &  5100  & 18.7  & 359  & 186 \\
           Phoenix-6 &  22.038990  & -52.749837  & 2018-08-07   & 58337.69  &  5700  & 18.7  & 348  & 249 \\
           Phoenix-7 &  20.661395  & -54.573383  & 2018-08-06   & 58336.70  &  7200  & 18.6  & 359  & 254 \\
              Ravi-0 & 344.584255  & -60.354707  & 2018-10-26   & 58417.50  &  6900  & 18.3  & 349  & 302 \\
              Ravi-1 & 343.013169  & -58.523030  & 2018-09-07   & 58368.51  &  7200  & 18.4  & 347  & 282 \\
              Ravi-2 & 341.596599  & -56.670068  & 2018-09-14   & 58375.50  &  7200  & 18.7  & 359  & 324 \\
              Ravi-3 & 340.313034  & -54.807171  & 2018-09-11   & 58372.49  &  7200  & 18.5  & 347  & 317 \\
              Ravi-4 & 339.142485  & -52.930510  & 2018-09-29   & 58390.52  &  7200  & 18.4  & 359  & 320 \\
              Ravi-7 & 336.166823  & -47.240314  & 2018-09-27   & 58388.50  &  7200  & 18.5  & 359  & 326 \\
              Ravi-8 & 335.318605  & -45.326721  & 2018-09-12   & 58373.43  &  7200  & 18.5  & 359  & 321 \\
       Turranburra-8 &  60.602100  & -18.782405  & 2018-10-24   & 58415.67  &  7200  & 18.4  & 349  & 269 \\
       Willka-Yaku-0 &  38.631197  & -57.506521  & 2018-09-16   & 58377.70  &  7600  & 18.6  & 347  & 239 \\
       Willka-Yaku-1 &  38.037160  & -59.482054  & 2018-09-11   & 58372.68  &  9600  & 18.7  & 347  & 223 \\
       Willka-Yaku-2 &  37.367524  & -61.453825  & 2018-09-12   & 58373.71  &  7200  & 18.7  & 347  & 234 \\
       Willka-Yaku-3 &  36.606647  & -63.423661  & 2018-09-13   & 58374.69  &  8100  & 18.8  & 347  & 285 \\
\hline
\end{tabular}
\end{center}
\end{table*}

\begin{table*}
    \contcaption{}
\begin{center}
\scriptsize
\begin{tabular}{l r r l l c c r r}
\hline
Field Name  & RA  & Decl & UTDATE & MJD & $t_\mathrm{exp}$ & r(S/N=5) & $N_\mathrm{targets}$ & $N_\mathrm{goodstar}$ \\
\hline
\hline
            Orphan-0 & 333.495287  & -35.792559  & 2018-09-16   & 58377.63  &  5400  & 18.0  &  96  &  67 \\
            Orphan-5 & 311.994544  & -65.291570  & 2018-09-14   & 58375.46  &  3000  & 17.8  & 347  & 316 \\
            Orphan-6 & 300.961865  & -70.170982  & 2018-10-24   & 58415.49  &  5400  & 17.3  & 347  & 318 \\
            Orphan-7 & 284.073327  & -73.980884  & 2018-10-24   & 58415.42  &  5400  & 17.5  & 336  & 324 \\
            Orphan-8 & 260.292330  & -75.842149  & 2018-09-14   & 58375.41  &  3000  & 17.9  & 359  & 336 \\
            Orphan-9 & 234.767419  & -75.015676  & 2018-09-16   & 58377.46  &  3600  & 17.6  & 348  & 336 \\
           Orphan-10 & 214.945014  & -71.858611  & 2018-09-16   & 58377.41  &  2700  & 17.5  & 359  & 337 \\
           Orphan-11 & 201.728540  & -67.327261  & 2018-09-14   & 58375.37  &  2400  & 18.1  & 347  & 316 \\
           Orphan-13 & 246.613855  & -75.778180  & 2018-09-29   & 58390.40  &  4500  & 17.8  & 359  & 336 \\
           Orphan-14 & 272.560733  & -75.241258  & 2018-09-27   & 58388.44  &  4500  & 18.3  & 349  & 332 \\
           Orphan-15 & 293.352126  & -72.317383  & 2018-09-29   & 58390.46  &  4500  & 18.1  & 349  & 336 \\
           Orphan-16 & 306.916290  & -67.777798  & 2018-10-25   & 58416.47  &  5400  & 17.9  & 359  & 333 \\
           Orphan-17 & 316.524807  & -62.124740  & 2018-10-26   & 58417.42  &  5400  & 18.1  & 355  & 318 \\
           Orphan-23 & 188.420550  & -58.979594  & 2019-04-04   & 58577.64  &  3600  & 18.2  & 354  & 330 \\
           Orphan-24 & 184.921533  & -55.456617  & 2019-04-01   & 58574.65  &  2700  & 17.4  & 354  & 344 \\
           Orphan-25 & 182.010398  & -51.845786  & 2019-04-05   & 58578.65  &  2700  & 17.2  & 352  & 346 \\
           Orphan-26 & 179.579995  & -48.163979  & 2019-04-03   & 58576.65  &  2700  & 17.2  & 354  & 329 \\
           Orphan-28 & 175.866616  & -40.612588  & 2019-04-02   & 58575.64  &  3600  & 17.7  & 354  & 297 \\
           Orphan-30 & 173.301936  & -32.863553  & 2019-04-04   & 58577.59  &  3600  & 18.0  & 342  & 318 \\
           Orphan-32 & 170.967451  & -25.124912  & 2019-04-02   & 58575.59  &  3600  & 18.0  & 343  & 304 \\
           Orphan-33 & 169.618942  & -21.320312  & 2019-04-06   & 58579.65  &  3060  & 17.5  & 354  & 249 \\
           Orphan-34 & 168.164652  & -17.552359  & 2019-04-03   & 58576.59  &  4500  & 18.3  & 313  & 266 \\
           Orphan-35 & 166.635130  & -13.805600  & 2019-04-06   & 58579.61  &  3300  & 18.3  & 350  & 269 \\
           Orphan-36 & 165.068830  & -10.070858  & 2019-04-05   & 58578.59  &  4500  & 18.5  & 312  & 246 \\
           Orphan-38 & 161.936058  &  -2.604385  & 2019-04-01   & 58574.58  &  4500  & 18.4  & 298  & 211 \\
              Pal5-1 & 224.228739  &  -5.306396  & 2018-08-02   & 58332.37  &  5400  & 18.9  & 348  & 325 \\
              Pal5-2 & 225.553103  &  -3.721754  & 2018-08-02   & 58332.43  &  5400  & 18.7  & 359  & 288 \\
              Pal5-5 & 237.497969  &   4.857110  & 2018-08-10   & 58340.36  &  7200  & 19.1  & 348  & 331 \\
              Pal5-6 & 240.053281  &   5.941191  & 2018-08-09   & 58339.37  &  7200  & 18.9  & 348  & 328 \\
              Sgr-m1 & 295.703295  & -34.788933  & 2018-08-12   & 58342.36  &  3600  & 17.6  & 344  & 342 \\
              Sgr-m2 & 286.648265  & -32.139498  & 2018-09-10   & 58371.37  &  1200  & 15.9  & 347  & 345 \\
              Sgr-m3 & 300.428337  & -35.849861  & 2018-10-26   & 58417.40  &  1800  & 17.6  & 315  & 309 \\
              Sgr-s1 & 297.638039  & -29.602086  & 2018-08-12   & 58342.41  &  4200  & 17.7  & 359  & 338 \\
              Sgr-s2 & 288.958424  & -27.174620  & 2018-09-12   & 58373.39  &  2600  & 17.7  & 347  & 342 \\
              Sgr-s3 & 301.946385  & -31.119809  & 2018-09-13   & 58374.37  &  3000  & 18.1  & 318  & 310 \\
\hline
       calib-NGC6316 & 259.155679  & -28.139205  & 2018-09-08   & 58369.37  &  1800  & 16.0  & 269  & 268 \\
       calib-NGC7078 & 322.492990  &  12.167678  & 2018-09-11   & 58372.38  &  1200  & 15.3  & 347  & 346 \\
       calib-NGC7089 & 323.364115  &  -0.823334  & 2018-09-16   & 58377.39  &   600  & 14.7  & 232  & 232 \\
          calib-Sgr1 & 283.900504  & -30.799532  & 2018-09-09   & 58370.36  &   600  & 16.4  & 347  & 347 \\
          calib-Sgr4 & 281.600246  & -30.100294  & 2018-09-07   & 58368.37  &  1800  & 15.1  & 347  & 347 \\
\hline
         atlas-test1 &  26.200215  & -30.599785  & 2018-01-23   & 58141.43  &  7200  & 18.5  & 360  & 284 \\
         atlas-test2 &  13.305955  & -23.062726  & 2018-06-06   & 58275.76  &  4500  & 18.6  & 360  & 260 \\
            Pal5-pt1 & 222.891667  &  -6.885655  & 2018-06-05   & 58274.58  &  7200  & 19.0  & 346  & 323 \\
            Pal5-pt2 & 222.959421  & -11.185382  & 2018-06-06   & 58275.36  &  7200  & 19.3  & 349  & 327 \\
          Jhelum-pt1 & 341.012565  & -50.501704  & 2018-06-06   & 58275.69  &  4800  & 17.9  & 349  & 344 \\
\hline
\end{tabular}
\end{center}

 \end{table*}

\subsection{Stream Targets (P9--P7)}\label{sec:stream_target}

We first cross-match DES DR1 with \gaia DR2 by selecting a DES DR1 nearest neighbour for each \gaia source having separation $< 1$\arcsec\ . We do not use the proper motion information to account for possible high proper motion stars in the cross-match because the DES observations were conducted mostly while the Gaia mission was ongoing (i.e., they were observed at the same epoch).  We then select our stellar targets from this joint catalogue as stream candidates. 

From this joint catalogue, we first perform a stellar selection when the objects have 
\begin{equation}
    \texttt{WAVG\_SPREAD\_MODEL\_I} < 0.005,
    \label{eq:sgsep}
\end{equation} 
or
\begin{equation}
   \texttt{ASTROMETRIC\_EXCESS\_NOISE} < 2,
\end{equation} 
where {\tt WAVG\_SPREAD\_MODEL\_I} is a weighted averaged (WAVG)
SExtractor model-based star-galaxy 
separation quantity \citep{Morganson2018} in $i-$band from DES DR1.
The {\tt ASTROMETRIC\_EXCESS\_NOISE} is the measure of the scatter of astrometric measurements around the solution from \gaia DR2 above what is expected from a noise model \citep{Lindegren2016,Lindegren2018}. This statistic identifies sources with bad astrometry and/or extended sources (e.g., galaxies) \citep{Koposov2017}. We note that this is a very conservative selection because we do not want to miss any possible stellar targets.
In addition, we reject stars with parallax measurements consistent with being local disk stars. Specifically, we perform a parallax cut of
\begin{equation}
    \text{\tt PARALLAX} - 3 \times \text{\tt PARALLAX\_ERROR} < 0.2,
    \label{eq:plxsel}
\end{equation} 
to remove stars with significant parallax measurements.

The bright-end magnitude limit is at $r_0\sim15$, which is close to the saturation limit of DES DR1.\footnote{We note that stars at $r_0\sim15$ may suffer some saturation problems. However, we still include these targets, as bright stream members are rare.} The faint end  magnitude limit is generally at $r_0\sim19.5$, but varies slightly from stream to stream. For example, considering the distance of the Elqui stream ($\gtrsim 40$~kpc), we set the faint end limit at $r_0=19.8$.  For closer streams, such as Jhelum and Indus, we set a brighter faint end at $r_0=19.0$.

\begin{figure*}
\centerline{\includegraphics[width=7in]{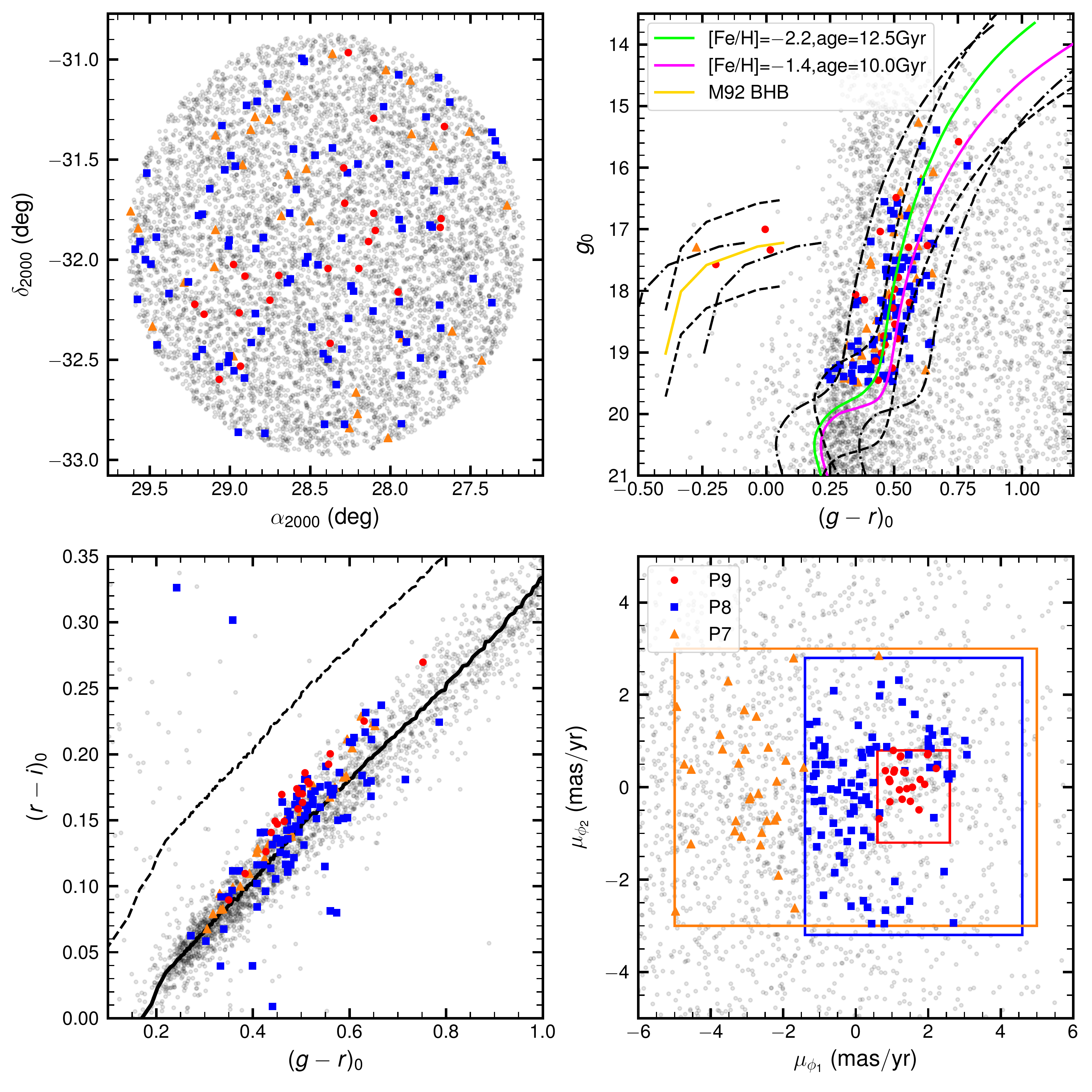}}
\caption{
Illustration of the stream target selection in one AAT field. For all panels, grey dots represents all stars in DES DR1 $\times$ \gaia DR2 in this field. P9, P8, P7 targets are shown in red circles, blue squares and orange triangles, respectively.  {\it Upper-left}: Spatial distribution on the sky of the stream targets in one AAT field. 
{\it Upper-right}: Stream targets selected in colour--magnitude space. A more metal-rich ($\feh = -1.4$, lime curve) and a metal-poor ($\feh = -2.2$, magenta curve) Dartmouth isochrone are used to guide the selection for giant and MSTO candidates. The M92 BHB ridgeline is used to guide the selection for BHB candidates. Both black dashed and dot-dashed lines show the bandwidth of the selection. 
{\it Lower-left}: Stream candidates in colour--colour space. The black solid line is the stellar locus in DES $(g-r)_{0}$ vs $(r-i)_{0}$,
and the dashed line is the stellar locus shifted by +0.06 mag in $(r-i)_{0}$. We select targets between these two lines as candidate metal-poor stars for the stream targets and used for P9 and P7 targets. 
{\it Lower-right}: Stream targets in proper motion space in stream coordinates ($\phi_1$, $\phi_2$). Proper motions shown here are all corrected for the Sun's reflex motion assuming stars are at the distance of the stream. Stream targets are selected to be centred on the proper motion of the stream measured in Shipp et al. (\textit{submitted\@}), with a tight PM cut for P9 targets (red box), a less tight PM cut for P8 targets (blue box), and a loose PM cut (orange box, independent of the detected stream PM) for P7 targets.
}
\label{fig:stream_target}
\end{figure*}

We then further subdivide the candidate stars using (i) isochrone filtering in colour--magnitude space; (ii) metal-poor star selection in colour--colour space; and (iii) likely member selection in proper motion (PM) space. Figure \ref{fig:stream_target} shows the selection process for one field in the ATLAS stream as an example. These selection criteria are as follows:

\begin{enumerate}
    \item {\it Colour--magnitude space (Figure \ref{fig:stream_target}, upper-right panel):} Considering the relatively metal-poor nature of known streams, we select targets in a window of either $|\Delta(g-r)| < 0.10$ or $|\Delta g| < 0.5$ from either a metal-poor ($\feh = -2.2$) or a relatively metal-rich ($\feh = -1.4$) Dartmouth isochrone \citep{Dotter2008} for red giant branch (RGB) and main sequence turnoff (MSTO) candidates. The same criteria are applied to select blue horizontal branch (BHB) candidates using a M92 BHB ridgeline \response{from \citet{Belokurov:2007}, built based on SDSS photometry from \citet{Clem2006PhDT}, and we transform from the SDSS photometry to the DES photometric system using equation (5) from \citet{Bechtol:2015}}. For some streams, when the target density is low, we also increase the bandwidth of the selection. We note that we purposely discard the red horizontal branch (RHB) candidates, given that the RHB has large contamination from the foreground MSTO stars, which would result in a lower member identification efficiency. 
    \item {\it Colour--colour space (Figure \ref{fig:stream_target}, lower-left panel ):} As shown in~\cite{Li:2018b} and \citet{Pace:2018}, the location of stars in a dereddened $g-r$ vs $r-i$ diagram is correlated with the metallicity of the star (discussed in Section \ref{sec:mpdiscuss}). Specifically, stars located above and to the left of the stellar locus (black solid line)
    tend to be more metal poor than those below and to the right of the locus. Therefore, we select targets in a band between the stellar locus and a locus shifted +0.06 mag in $r-i$ (the black dashed line) as the metal-poor targets. 
    \item {\it Proper motion space (Figure \ref{fig:stream_target}, lower-right panel):} \gaia DR2 proper motions greatly improve our target selection efficiency. The proper motion of each DES stream is measured in Shipp et al. (\textit{submitted\@}). \SSSSS target selection used a preliminary version of these proper motions.
%    \FIXME{Do we need to list the prelim version we used in the appendix of this paper for completeness?}
    For a given stream, three PM categories are selected: 
    \begin{itemize}
	\item PM1: a tight PM selection with $|\mui - \muinot| < 1 \masyr$;
	\item PM2: a less tight PM selection with $|\mui - \muinot| < 2\sim3 \masyr$  (varying from stream to stream);
	\item PM3: a loose PM selection with $|\muone| < 4\sim5 \masyr$ and $|\mutwo| < 2\sim3 \masyr$ (varying from stream to stream);
    \end{itemize}
    where $i$ = 1 or 2 and  \muonenot and \mutwonot\ are the PM of the stream and \muone and \mutwo are the PM in stream coordinates of the target star after solar reflex motion correction (assuming all the targets are at the distance of the stream from isochrone fitting).
\end{enumerate}

We then assign stream targets to priority levels P9, P8, and P7, as presented in Table \ref{table:targets}. All three categories have the same selection in colour--magnitude space.
P9 targets satisfy both the metal-poor and PM1 selection because halo stream members are largely metal-poor (this will be further demonstrated in future \SSSSS papers). P8 targets have PM2 selection, with targets in P9 excluded. P9 targets are essentially a subset of P8 targets that is given higher priority in the case of fibre collisions. P7 targets meet the metal-poor and PM3 selection criteria (with P9 and P8 targets being excluded). Note that the PM3 selection is independent of the measured proper motion of the stream. This is to ensure that in the case that the proper motion of a stream was measured incorrectly, our target selection would still include some of the stream members. We choose a smaller range in \mutwo because the transverse motion of the stream (after solar reflex motion correction) is expected to be small except for cases where the streams have suffered large gravitational perturbations \citep[see, e.g., the Orphan Stream;][]{Koposov:2019, Erkal:2019b}.

The number density of targets per pointing varies from stream to stream, mainly dependent on the Galactic latitude of the field, ranging from $\sim$(20, 90, 30) stars in (P9, P8, P7) for the ATLAS stream (at $b\sim-80\degr$) to $\sim$(90, 200, 50) for the Chenab stream (at $b\sim-40\degr$). To achieve the best fibre efficiency, we also vary the bandwidth of the isochrone filtering and the proper motion selection, as described above. We note that these systematic selection criteria (i.e. selection in parallax, colour, magnitude and proper motion simultaneously) yield a factor of 20--100$\times$ reduction in target density, mostly eliminating the foreground contamination. 

No additional spatial selection is performed within the AAT fields. In other words, all targets that pass the criteria described above in one AAT field are sent to \code{configure} to be assigned fibres according to their priority. We note that some of the streams are much narrower than the FOV of 2dF (e.g. the width of the ATLAS stream is 0.25\degr). We treat all targets within one AAT field equally, allowing us to explore possible variations in stream width, as well as the possibility of a non-Gaussian density profile across the streams.

\subsection{Other Stellar Targets (P6--P3)}\label{sec:other_targets}

Thanks to the efficient target selection described in Section \ref{sec:stream_target} and the high multiplex capability of 2dF (i.e. 392 science fibres), we are able to use spare fibres for a Milky Way halo star survey and a low-z galaxy survey (Section \ref{sec:galaxy_targets}) in the stream fields, especially in the fields at high Galactic latitude. Due to the limited number of fibres, neither the halo survey nor the low-$z$ survey is designed to be complete or uniform. 

Our target selection for the halo survey has a complicated selection function as shown below. Scientifically, we intend to use the limited number of spare fibres to find interesting objects such as hyper-velocity stars, extremely metal-poor stars, moving groups in the halo, etc. We note that a reconstruction of the survey selection function for the halo might be difficult and was not a goal when we designed this auxiliary survey.

For the Milky Way stellar halo targets, we first perform the same stellar selection and parallax selection as described in Eq.~\ref{eq:sgsep} -- \ref{eq:plxsel} in Section \ref{sec:stream_target}, except for the nearby white dwarf targets, which are described in Section \ref{sec:wd}. We then select stars meeting various criteria and assign them to the P6-P3 categories. Specifically, P6 stars are the highest priority among all non-stream targets, and are composed of several rare object types as described in Section \ref{sec:bhb} to Section \ref{sec:wd}. We note again that when the targets are selected in multiple priority categories, the highest priority is used as the input to the fibre allocation software.

\subsubsection{P6: Blue Stars}\label{sec:bhb}

Since blue stars are generally rare and bright, we set P6 for blue stars with $-0.4<(g-r)_0<0.1$ and $15 < g_0 \lesssim 19.5$, where the faint end limit varies from stream to stream. Most stars in this selection are either blue horizontal branch stars (BHBs) or blue stragglers (BSs). The selection results in, on average, about 20-50 blue stellar targets within an AAT field, although it turns out that about one-third of these blue stellar targets are actually QSOs (see Section \ref{sec:qsos}).

\subsubsection{P6: RR Lyrae stars}

RR Lyrae candidate stars are selected from two separate source catalogues, table \texttt{vari\_classifier\_result} and table \texttt{vari\_rrlyrae}, released as part of the \gaia DR2 
\citep[see][]{Clementini2018,Holl2018}. The astrometry and photometry information are acquired by joining with the main \texttt{gaia\_source} catalogue. We then selected the RR Lyrae targets with $15 < G < 20$. This results in several (1--5) stars on average per AAT field.

\subsubsection{P6: White Dwarfs}\label{sec:wd}

We also include hot white dwarf candidates (WDs) in the P6 category. 
We note that these WDs are not necessarily halo stars, but they are considered part of the halo survey due to their low target density.  Our interest in including these hot WD candidates as targets is in their potential future use as faint spectrophotometric standard stars for large surveys and large instruments in the Southern Hemisphere \citep[e.g.,][]{Narayan2019}.

Candidate hot WDs were selected from \gaia DR2, based on criteria for identifying WDs from the \gaia DR2 photometry and astrometry as described in \citet{GentileFusillo:2019}. When we created our sample, their paper was still in preprint form and we did not have a copy of their catalogue of candidate WDs, so we applied their criteria (with minor variations) to regenerate their final catalogue 
%of 486,641 candidate WDs 
in the \gaia DR2 data ourselves.
We further trimmed our sample using the following prescription:
\begin{itemize}
    \item Since we are primarily interested in hot WDs within the DES footprint, we matched our catalogue to the DES DR1 catalogue, removing entries that had no matches.
    \item We also used the $T_{\rm eff}$ values from Sloan Digital Sky Survey \citep[SDSS;][]{York:2000} stars modelled by the SEGUE Stellar Parameter Pipeline \citep{Lee:2008} to identify a colour cut that would select only those candidate WDs with $T_{\rm eff} \gtrsim 10,000$~K.  We then applied that colour cut ($g-r \lesssim 0.0$) to the WD candidates remaining from our match with the DES catalogue.
    \item In order to not waste fibres on candidates that had a low probability of being actual WDs, we imposed cuts based on \gaia photometry and astrometry that would include only those candidates with a probability of being a white dwarf of $P_{\rm WD} \gtrsim 0.80$ from \citet{GentileFusillo:2019}.
    \item To avoid unnecessary duplication, we also excluded any white dwarf candidate that already had a spectrum from SDSS.
\end{itemize}

\begin{figure}
\centerline{\includegraphics[width=3.5in]{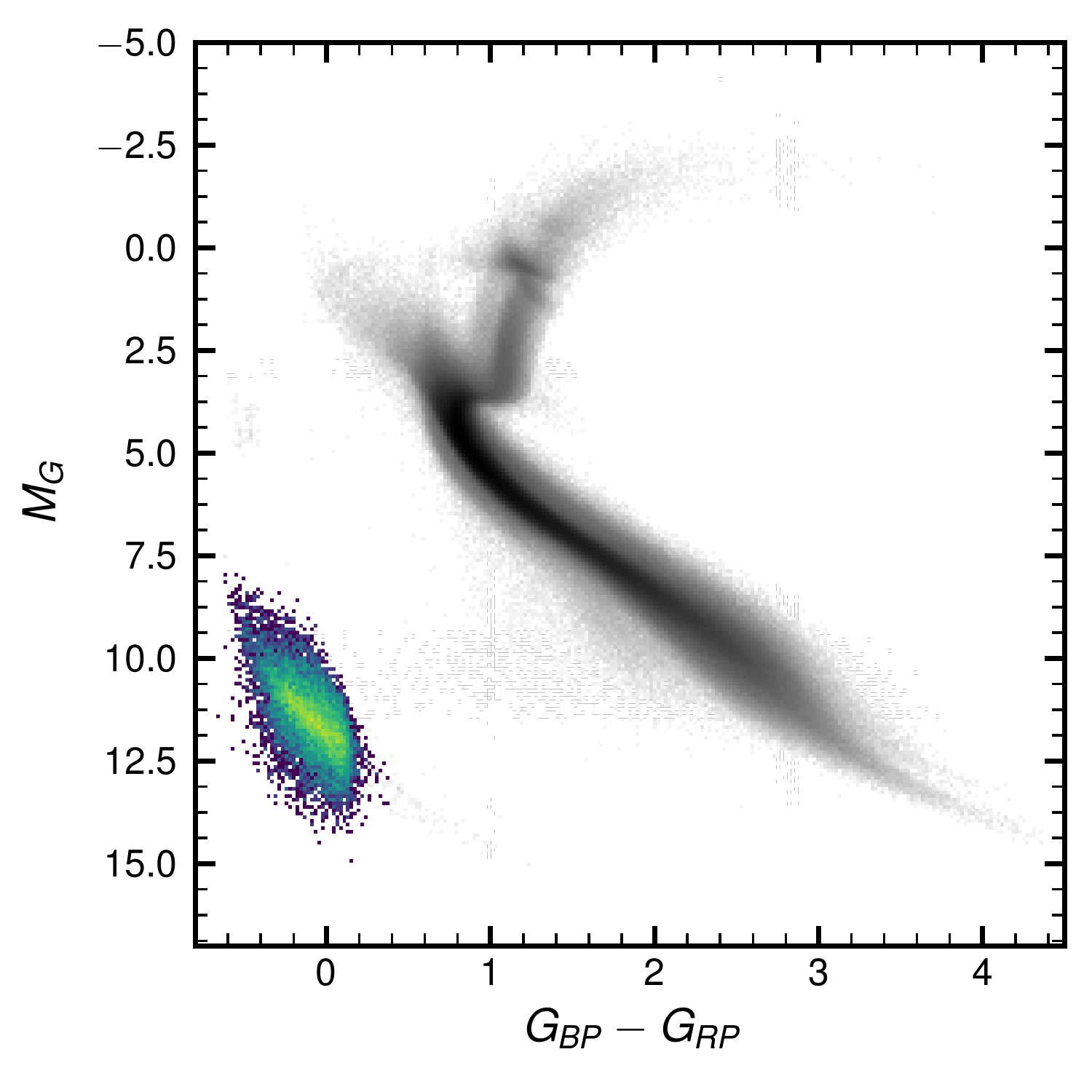}}
\caption{
\gaia HR diagram constructed with all Gaia stars at {\tt PARALLAX}/{\tt PARALLAX\_ERROR} $>10$ and $E(B-V) < 0.02$. Also plotted as a 2D histogram is our targeted hot white dwarf candidate sample in the lower left corner, with yellow indicating higher density in each bin.  
}
\label{fig:gaia_dr2_wds_hr_diagram}
\end{figure}

A plot showing our candidate hot WD targets in the \gaia HR diagram can be found in Figure~\ref{fig:gaia_dr2_wds_hr_diagram}.
There are 13,019 candidate hot WD candidates over the full DES footprint in our list of potential P6 targets, and typically a few (1--4) were observed in each AAT field.

\subsubsection{P5: Extremely Metal-Poor Candidates}\label{sec:emp}

We target, with priority P5, candidate stars selected as part of the SkyMapper search for extremely metal-poor (EMP) stars. As described by Da Costa et al.\@(\textit{submitted}), these stars are selected using $vgi$ photometry from Data Release 1.1 of the SkyMapper Southern Survey \citep{Wolf:18}. While the SkyMapper EMP program usually imposes a faint limit of $g\approx16$, we relax this to $g=17.5$ to boost the number of candidates per 2dF field to typically in the range $\sim5-10$. Unsurprisingly, duplicate entries in the SkyMapper EMP list and the DES metal-poor halo star list (Section \ref{sec:mp}) sometimes occur; since the SkyMapper targets have P5 while the DES targets have P4, objects are preferentially allocated from the SkyMapper list.

\subsubsection{P4: Metal-Poor Stars}\label{sec:mp}

P4 targets are selected to be metal-poor candidates using the dereddened $g-r$ vs $r-i$ colour of the stars, in a similar way as metal-poor star selection for streams described in Section \ref{sec:stream_target} and in the lower left panel of Figure \ref{fig:stream_target}. To further minimise the target density, 
we select metal-poor targets that lie between 0.02 and 0.06 mag in $r - i$ above the empirical stellar locus, 
and $0.4 < (g-r)_0 < 1.0$, $15 < g_0 < 18.5$. This selection results in an additional  $\sim 10-50$ targets per AAT field. For stream fields at low Galactic latitude ($|b|\lesssim 50\degr$), P4 targets are not selected.

\subsubsection{P3: Low Proper Motion Stars}

P3 targets are selected to be stars with small proper motion and therefore are more likely to be distant halo stars. To make this selection, we first compute a reflex motion corrected proper motion for each star, based on their position on the sky, assuming that they are all at 30~kpc from the Sun. We then select the targets with $|\mu_\alpha| < 3\masyr$ and $|\mu_\delta| < 3\masyr$ and $15<g_0<18.5$. This selection results in $\sim 10-50$ targets per AAT field (depending on the Galactic latitude of each field). For stream fields at lower Galactic latitudes ($|b|\lesssim  50\degr$), P3 targets are not observed.

\subsection{Low-$z$ Galaxy Targets (P2--P1)}\label{sec:galaxy_targets}

Observations of nearby dwarf galaxies ($z < 0.02$, $M_r > -16$) are critical for understanding the mapping between dark matter and galaxy formation \citep{Geha:2017}.
However, these galaxies are difficult to distinguish from the far more numerous background galaxy population via photometry alone.  The goal of including low-redshift (low-$z$) galaxy targets in \SSSSS is to increase the number of spectroscopically confirmed low-$z$ galaxies in order to better train photometric selection algorithms, and help build a statistical sample of very low-$z$ galaxies. 

The galaxy targets are selected using the DES DR1 catalogue and are limited to the DES stream fields. To build the galaxy target list, we first select the objects that satisfy all of the following conditions:
\begin{align*}
 & \text{\tt IMAFLAGS\_ISO\_R} = 0,  \\
 & \text{\tt FLAGS\_R} < 4, and\\
 & \text{\tt EXTENDED\_COADD} = 3,
\end{align*}
where the first two criteria are to select clean objects and {\tt EXTENDED\_COADD} is defined as: 
\begin{align*}
& {\tt EXTENDED\_COADD} = \\
& (\text{\tt SPREAD\_MODEL\_R}+ 3 \times \text{\tt SPREADERR\_MODEL\_R} > 0.005) \\
 + & (\text{\tt SPREAD\_MODEL\_R}+\text{\tt SPREADERR\_MODEL\_R} > 0.003) \\
 + & (\text{\tt SPREAD\_MODEL\_R}-\text{\tt SPREADERR\_MODEL\_R} > 0.003),
\end{align*}
to select high-confidence galaxies based on SExtractor model-based star-galaxy separation\footnote{\url{https://des.ncsa.illinois.edu/releases/dr1/dr1-faq\#faq1}}. 

We also limit the galaxy targets to the magnitude range of $18 < r_0 < 20$, and to the fields within the Galactic Cap ($|b| > 50\degr$).

\begin{figure*}
\centering
\includegraphics[width=0.99\textwidth,clip,trim=0 0.5cm 0.2cm 0.4cm]{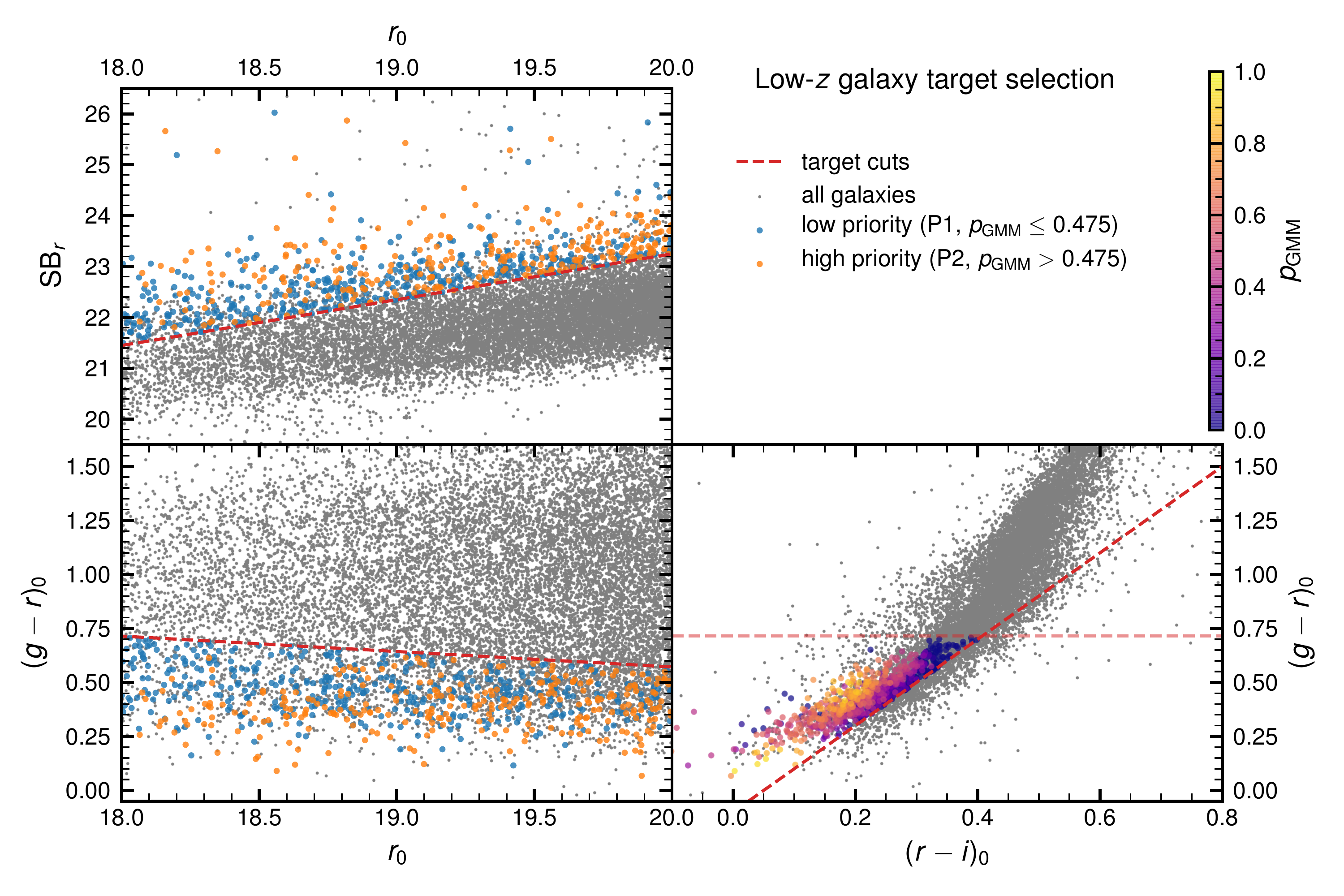}
\caption{The three panels demonstrate our low-$z$ galaxy target selection in the surface brightness--magnitude (\textit{upper left}), colour--magnitude (\textit{lower left}), and colour--colour (\textit{lower-right}) spaces. 
The red dashed lines in each panel show the photometric cuts that preferentially select the very low-$z$ galaxies (see Sec.~\ref{sec:galaxy_targets} for cut definitions), and we select only galaxies that pass all three cuts to be our targets. 
We then use a multivariate Gaussian Mixture Model in the $grizY$ space to assign a probability ($p_\text{GMM}$) to each of the galaxy candidates (as shown by the colour dots in the lower right panel), and assign candidates that have $p_\text{GMM} > 0.475$ to P2 (high-priority low-$z$ targets, shown as orange dots in the left panels), and the rest to P1 (low-priority low-$z$ targets, shown as blue dots in the left panels).
}
\label{fig:lowz-target-sel}
\end{figure*}

After the initial selection, we then use the low-$z$ galaxy data from the SAGA Survey\footnote{\url{http://sagasurvey.org}} \citep{Geha:2017} and the method outlined in Mao et al.\@ (\textit{in prep.\@}) to develop a set of photometric cuts that preferentially select very low-$z$ galaxy candidates.
They are cuts in the colour--colour, colour--magnitude, and surface brightness--magnitude spaces:
\begin{align*}
& (g_0-r_0) > (r_0- i_0 - 0.05) \times 2 ; \\
& (g_0-r_0) < 2 - (r_0 / 14); \\
& \text{SB}_r > 0.9 r_0 + 5.25;
\end{align*}
and are shown in Figure~\ref{fig:lowz-target-sel}. Here, $\text{SB}_r$ is the surface brightness derived from $r$-band magnitude and flux radius. We only select galaxies that pass all three photometric cuts. 
We then further prioritize these candidates into high (P2) and low (P1) priority using a multivariate Gaussian Mixture Model (GMM) trained in colour space ($grizY$) on both synthetic data and SAGA spectroscopic data. The GMM probabilities are also shown in the colour--colour panel of Figure~\ref{fig:lowz-target-sel} for reference. 

While our photometric cuts preserve very high completeness for very low-$z$ (up to $z<0.02$) galaxies (Mao et al.\@ \textit{in prep.\@}), due to incomplete sampling of these lower priority targets, the resulting sample cannot be considered complete. However, even an incomplete sample serves our goal of obtaining training data for photometric selection algorithms that are tuned to very low-$z$ ranges.

\subsection{Stream Overlaps}

Some AAT fields targeting different streams have partial overlap. In these cases, we have only observed the fields defined for one of the streams, but we select stream candidates from both streams as targets. These fields include (see Figure \ref{fig:s5pointing} and Table \ref{table:fields}) Chenab-3 (overlap w/ Jhelum), Chenab-5 (w/ Ravi), Chenab-6 (w/ Ravi), Jhelum-9 (w/ Indus), Jhelum-10 (w/ Indus). In these fields, P9 and P7 are the two categories for stream candidates in the primary stream, and P8 and P6 are the two categories for the stream candidates in the overlapping stream. No further targets from the halo or low-z surveys are considered in these fields. 

\subsection{Streams Beyond DES}\label{sec:nonDES}

While this paper focuses on the target selection and observations for the DES streams, we also observed some streams beyond the DES footprint, including the Orphan stream, the Sagittarius stream, and the Palomar 5 stream. Observations of these streams were taken while the DES streams were not observable for parts of the night. We used different input photometric catalogues, target selection and field selection criteria for each stream, to fulfill different science goals on each stream. For example, Orphan stream targets are selected to be the extensions of the Chenab stream using \gaia DR2 photometry and proper motion in order to map the entire Orphan Stream in the Southern Hemisphere. Sagittarius stream targets were selected to study  the stream bifurcation using \gaia DR2  photometry and proper motion. Palomar 5 stream candidates were targeted using \response{Pan-STARRS1 photometry~\citep{Chambers2016}} and \gaia proper motions to search the extension beyond the known length of the stream. Since each stream was treated differently, we will leave a more detailed description of the data on these streams for future publications. We note, however, that the data collected for these streams were reduced and validated alongside the rest of the \SSSSS data, as discussed in Section \ref{sec:observations} and \ref{sec:validation}.

We also note that in 2019, \SSSSS plans to extend the survey beyond the DES streams and map more streams at $\delta_{2000} < 30\degr$. 

\section{Observations and Reduction}\label{sec:observations}

\subsection{Observations}
As previously noted, \SSSSS\ used the AAOmega spectrograph on the 3.9-m Anglo-Australian Telescope, located at the Siding Spring Observatory in Australia. AAOmega is a dual arm spectrograph, with the light split by a dichroic centred at 5800~\AA. The gratings employed were 580V on the blue arm, and 1700D on the red arm, corresponding to spectral resolutions of $\sim1300$ and $\sim 10000$. With these, the blue side wavelength coverage is $3800-5800$ \AA, while the coverage on the red side is $8420-8820$ \AA.  The gratings were chosen so that we could have the highest spectral resolution in the red centred on the near-infrared calcium triplet (CaT) lines to derive precise radial velocities of stream members, and the largest spectral coverage in the blue for fainter stars as well as galaxies for spectroscopic redshift determination.

To obtain sufficient signal-to-noise (S/N) on our faintest targets, each DES stream field was observed with a total integration time of $\sim$7200 seconds,\footnote{For non-DES streams, the integration time varies from stream to stream depending on the science goals.} split into three equal exposures to mitigate cosmic ray contamination.
The average resulting S/N of stellar targets at $r\sim18.5-19.0$ (see Table \ref{table:fields})
is $\sim5$ per pixel in the red arm (at a pixel scale of $\sim0.23$~\AA~pixel$^{-1}$), allowing velocity determinations at a precision of $\sim 1$~km~s$^{-1}$.
Furthermore, calibration exposures, consisting of arc spectra and a quartz fibre flat field, were obtained for each field right before or after the science exposures, while a series of bias exposures were obtained before the night's observing began.

The observation date and exposure time for each field are listed in Table \ref{table:fields}. We re-observed a few fields if the first observation on the field was obtained in unfavourable weather conditions; in such cases, Table \ref{table:fields} only includes the observations taken under the best conditions.

We had a total of 25 nights of observing time spread over 29 nights (with some half-nights) spanning from August 2018 to October 2018. We lost approximately 5 nights in total due to poor weather (either too cloudy or seeing > 3\arcsec). The remaining 20 nights had good weather, with an average seeing of $\sim1\farcs5$. In April 2019, we obtained another $\sim12$ hr of observations with good seeing conditions of $\sim1\farcs5$ or better, which we devoted entirely to the Orphan Stream because the DES streams were not visible. More observations are planned and will be executed later in 2019.

During our observations, we found that some fibres have lower-than-expected throughput, likely caused by the fibre placement accuracy. This can severely degrade the S/N for these fibres especially under good seeing conditions (seeing $< 1$\arcsec), and we discuss this issue in more detail in Appendix \ref{app:pointing}.

\subsection{Data Reduction}\label{sec:reduction}

\subsubsection{\code{2dfdr} reduction}
\label{sec:2dfdr}

The initial data reduction was undertaken with the \code{2dfdr} software package \citep{2015ascl.soft05015A}, which automatically performs the standard reduction steps for multi-fibre data: debiasing the CCD frames, tracing the location of the stellar spectra from the location of tramlines drawn from the fibre flat, then wavelength calibration, and extracting the 1D spectra.

The blue arm (580V grating) data were reduced using the OzDES~\citep{Yuan2015} reduction parameter files. The red arm (1700D grating) data were reduced using the default settings, except for the following changes: we chose a 2D fit for the scattered light subtraction, a $7^{\rm th}$ order polynomial fit for the fitting of the wavelength solution of the arcs, and a $1^{\rm st}$ order polynomial fit to the sky lines for additional wavelength calibration. 

We note that one-quarter of the observations were taken at or near full-moon. For those observations, the extracted 1D spectra from the blue arm data show  negative fluxes in the continuum or contamination by solar spectrum, which was likely caused by imperfect sky subtraction when the sky background is strong. Therefore, the blue arm spectra taken under full moon should be used with caution. For stellar targets, as discussed later in Section \ref{sec:validation}, we mostly used the measurements from the red arm spectra for future analysis. For galaxy targets, since only blue arm spectra were used for redshift determination (see Section \ref{sec:galaxy-z}), those spectra suffering strong sky background were discarded.

\subsubsection{Fitting the spectra with \code{rvspecfit}}\label{sec:rvspecfit}

To determine the spectral atmospheric parameters and radial velocities (RVs) of each star, we have run each targeted spectrum through the template fitting code \code{rvspecfit}\footnote{\url{https://github.com/segasai/rvspecfit}} built for large stellar survey RV fitting. The code is loosely built on the template fitting described in \citet{Koposov2011}. 
Given the stellar template $T(\lambda|\boldsymbol{\phi}, v)$, stellar atmospheric parameters $\boldsymbol{\phi}$, radial velocity $v$, and observed spectra ${D_i}$ with errors ${E_i}$ observed at wavelengths $\lambda_i$ at pixels $i$, the code performs a least-squares fit to the observed spectra using a spectral template multiplied by a polynomial continuum  $T(\lambda|\boldsymbol{\phi},v)(\sum_j a_j \lambda^j)$. 
Thus \code{rvspecfit} provides the log-likelihood of the data given stellar atmospheric parameters after marginalizing over polynomial continuum coefficients. 
The stellar templates are determined by the following stellar atmospheric parameters $\boldsymbol{\phi}$: effective temperature \teff, surface gravity \logg, metallicity \feh and alpha elements abundance \alphafe. 
For a given set of stellar atmospheric parameters, a stellar template is generated through a two-stage interpolation procedure. First we take the PHOENIX-2.0 high-resolution stellar spectra library~\citep{Husser2013} ,\footnote{\url{http://phoenix.astro.physik.uni-goettingen.de/}} which have been computed on a sparse grid of stellar atmospheric parameters. We note that the step-size of the grid is quite large ($\Delta \logg=0.5$, $\Delta \feh = 0.5$ to 1). We truncate the spectra in the grid to the AAT wavelength range and convolve them to the appropriate resolution ($R\sim1300$ for 580V and $R\sim10000$ for 1700D). After that we use the Radial Basis Function (RBF) multiquadric interpolation over the grid to evaluate templates on a stellar atmospheric parameter grid with smaller and uniform steps in \feh ($0.25$\,dex) and \alphafe grid ($0.2$ dex), while preserving the uniform step of $0.5$\,dex in \logg and non-uniform sampling of \teff from the original grid. This creates a finer, more uniform grid and fills in some isolated gaps present in the original PHOENIX-2.0 grid. The multiquadric interpolation step is only performed once when preparing for fitting of the AAT instrument spectra. A final stage of stellar template generation is performed during each likelihood evaluation on each observed spectrum from \code{2dfdr}.  It is done by \code{rvspecfit} code using linear N-D interpolation between the templates based on the Delaunay triangulation~\citep[see e.g.][]{Amidror2002} as implemented in \code{scipy.interpolate.LinearNDInterpolate}. This interpolation is fast enough to be done in each likelihood evaluation and provides smoothly changing spectral templates as a function of stellar atmospheric parameters.
 
With the data likelihood function described above, we sample the posterior of stellar atmospheric parameters and radial velocities of each star. To initialise the starting points of the Markov Chain, the fits are preceded by a cross-correlation step over a subset of templates,  followed by a Nelder-Mead search of the maximum likelihood point in the space of stellar atmospheric parameters and RVs. 

The priors adopted for the MCMC sampling are uniform over \logg, \feh, \alphafe, and RV. The prior range for the stellar parameters are determined by PHOENIX-2.0 limits; for RV we set the range to be between $\pm 2000$\,\kms.  The only informative prior used is on the effective temperature \teff, for which the prior is based on the colours and metallicities of the stars.  We have two separate \teff prior models, 
$\pi(\teff| {Gaia \rm\,photometry})$ and $\pi(\teff| { \rm DECam\,photometry})$. The latter is used in the fitting when DECam photometry is available, from either DES or DECaLS~\citep{Dey:2018}, and the former is used when only Gaia photometry is available. 
Rather than trying to construct a conventional  polynomial prior for \teff based on colours and metallicity \citep{Alonso99}, we fit a function $\log T_\mathrm{eff}(\mathrm{colours}, \feh)$ using a gradient-boosted tree \citep[see e.g.][]{Bishop2006} as implemented in \code{sklearn.ensemble}. We use the SDSS and SEGUE effective temperatures from SDSS DR9  \citep{Lee:2008, AllendePrieto2008} and the $G_{BP}-G_{RP}$ Gaia colours and $g-r$,\,$r-z$ DECam colours to train the model.\footnote{$i$-band photometry is not used because no $i$-band observations were taken by DECaLS.} 

Specifically we fit three functions,  $T_{\mathrm{eff},50}(\mathrm{colours}, \feh)$, $T_{\mathrm{eff},16}(\mathrm{colours}, \feh)$, $T_{\mathrm{eff},84}(\mathrm{colours}, \feh)$,
using quantile regression corresponding to the 16\%, 50\%, 84\% percentiles of the $\teff|\mathrm{colours}, \feh$ distribution, which we then use to define a log-normal prior on the effective temperature:
\begin{align}
{\mathcal P}(\log T_{\mathrm{eff}}|\mathrm{colours},\feh) &=  {\mathcal N}\Big(\log T_{\mathrm{eff}} | \nonumber \\ 
& \log T_{\mathrm{eff},50}\,,\,\frac{1}{2} (\log T_{\mathrm{eff},84} - \log T_{\mathrm{eff},16})\Big)\nonumber
\end{align}
conditional on the star's colour and \feh. 

\begin{figure*}
    \centering
    \includegraphics[width=7in]{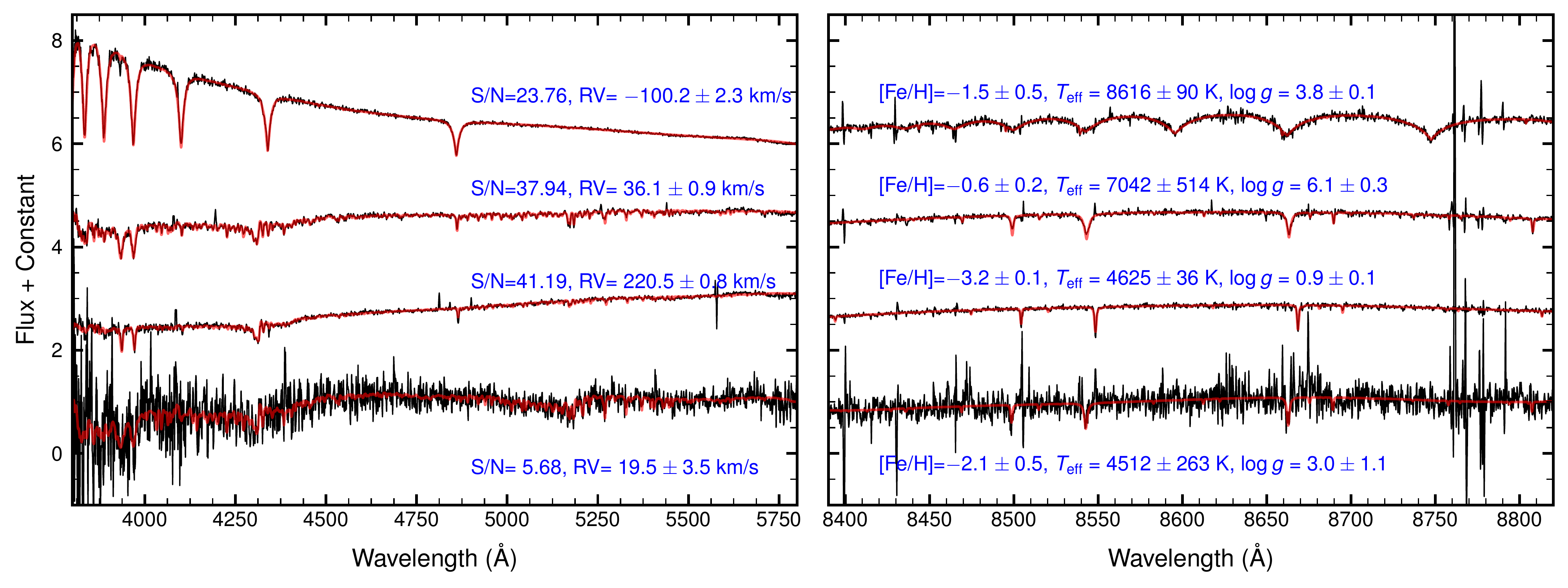}
    \caption{Examples of reduced 1D spectra from the blue arm ({\it left}) and red arm ({\it right}), spanning a range of S/N, \feh, \teff, and \logg (black lines), over-plotted with the best fit model templates from \code{rvspecfit} (red lines). Best-fit parameters and uncertainties from the red arm spectra are shown. The y-axis represents the measured flux plus a constant offset for ease of visualization. }
    \label{fig:spectra}
\end{figure*}

The posterior on \teff, \logg, \feh, \alphafe and radial velocity was sampled for each star using the ensemble sampler \code{emcee} \citep{Goodman2010,Foreman_Mackey:2013} with 60 walkers for at least 2000 iterations. The first 1000 iterations of the chain were treated as burn-in and were discarded from the final posterior distribution. We verify the chain convergence by computing the Geweke scores \citep{Geweke92} on each parameter and continue sampling until a satisfactory score is reached. We then use the chain to compute the best fit stellar atmospheric parameters and RVs. For most parameters we use and report the median and standard deviation  from the posterior chains. The measured quantities and their uncertainties are validated in Section \ref{sec:validation}. As during the validation we observe that the uncertainties on the RV  and \feh are somewhat underestimated, we adjust them according to the validation results (see Sections ~\ref{sec:rv_validation},\ref{sec:feh_validation}). In Figure \ref{fig:spectra}, we show examples of reduced 1D spectra together with the best-fit model templates.
 
Currently we fit the blue arm and red arm spectra with \code{rvspecfit} independently from each other.
The results from the red arm spectra are used for most of the analysis work in this paper and will likely be the basis for the future \SSSSS science papers.  Furthermore, except for studying the repeatability of the measurements (e.g. in Section \ref{sec:rv_validation}), we usually use the values from the spectrum with the highest S/N, when multiple observations were taken on a given object.

\subsubsection{CaT Metallicity}\label{sec:cat}

In addition to \code{rvspecfit}, we determined the metallicities using the equivalent widths (EWs) of the CaT lines from the red arm spectra. This is an independent check on the metallicity measurements for the RGB stream members.
We fit all three of the CaT lines with a Gaussian plus Lorentzian function. We then converted the summed EWs of the three CaT lines to \feh using the calibration relation as a function of absolute V magnitude from~\citet{carrera13}.\footnote{We transformed from DES-$g,r$ to V mag using equation (2) in~\citet{Bechtol:2015}.} 
In order to derive the absolute magnitude of each star, the distance to the star is needed. Therefore, the CaT metallicity derived here are only valid for stream members where the distance to the stream is known.
The uncertainties on the EWs are calculated uncertainties from the Gaussian and Lorentzian fit plus a systematic uncertainty of 0.2~\AA\ added in quadrature. This systematic uncertainty is derived by checking the EWs from the repeated measurements \citep{eri2, Li2018}, in a similar way as described in Section \ref{sec:rv_validation}. The metallicity uncertainties are calculated from the uncertainties on the CaT EWs and the uncertainties on the calibration parameters from~\citet{carrera13}. Note that we do not include any uncertainty from the distance to the stars. Although distance uncertainties are usually reported with the paper announcing the discovery of the stream, a distance gradient is usually not initially determined, though it is present in most streams. A shift of 0.3 mag in distance modulus will cause a change in derived CaT metallicity of $\sim0.05$~dex.

We note that the metallicity calibration relation from \citet{carrera13} only applies to RGB members and therefore the CaT metallicity derived here does not apply to stream members not on the RGB or to stream non-members.

\subsubsection{Galaxy Redshifts}
\label{sec:galaxy-z}

We independently determined redshifts of all blue arm spectra using \code{autoz} \citep{Baldry:2014}.  While \code{autoz} can in principle provide the redshifts for all the stellar objects, it mainly focuses on determining accurate extragalactic redshifts. Therefore we only used the results from \code{autoz} on non-stellar objects. All redshifts were visually inspected using \code{marz} \citep{Marz:2016}.   Among the $\sim3000$ targeted galaxies, $\sim2300$ of them were observed when the moon is less bright and therefore have robust redshift measurements. 

We found that a non-negligible fraction ($\sim4$\%) of our stellar targets turn out to be QSOs based on the presence of broad emission lines. The QSO redshifts were measured using \code{autoz}.  Secure redshifts for 674 QSOs are presented in Table \ref{table:qso} of Appendix B.  An additional 412 QSOs candidates were identified, but the limited spectral coverage included only a single broad emission line that could not be unambiguously identified.  As QSOs are contaminants to our stellar sample, we removed QSOs using a photometric selection described in Section \ref{sec:qsos}.

\section{Survey Validation and Quality Assurance}\label{sec:validation}

In order to assess the measurement quality of the \SSSSS pipeline, we observed several calibration fields during evening and morning twilight of the 2018 observing runs. These fields include a few globular clusters with metallicities ranging from $-$2.5 to $-$0.5, and fields in the Sagittarius stream (see Table \ref{table:fields}); targets in each field were selected from APOGEE \citep[SDSS DR14;][]{Majewski:2017,Abolfathi2018} with magnitude range $12 < G < 16$. Since the targets are bright, the exposure time is less than $30$ minutes for each field. The spectra were reduced and fit using exactly the same pipeline as described in Section \ref{sec:2dfdr} and \ref{sec:rvspecfit}. We then compared our derived parameters to the reported values to assess their accuracy. On top of dedicated APOGEE observations as validation we also use the measurements  from LAMOST DR4 \citep{Cui2012}, Gaia-ESO Survey (GES) DR3 \citep{Gilmore2012}, SDSS/SEGUE \citep{AllendePrieto2008} and GALAH DR2.1 \citep{Buder:2018} for stars from each survey that were serendipitously observed by \SSSSS. As the main science goals of \SSSSS are the stellar streams and Milky Way halo, we are mostly interested in the RVs and \feh measurements, thus we will focus on validating those two parameters in this section.

We note that the RVs and metallicities are derived independently from the blue arm and red arm spectra. For RVs, it is clear that the higher spectral resolution of the red arm should provide much better velocity precision for all but the bluest objects. For metallicities however, due to the much larger number of lines in the blue, as opposed to mostly CaT lines in the red, we expect the blue arm to be very competitive in abundance precision. However, we found that the red arm provides smaller systematic errors on metallicities at a cost of somewhat larger scatter. Therefore for the rest of the paper we mostly focus on the measurements from the red arm spectra for both RVs and metallicities. 
We may also use results from the blue arm spectra in the future, as they may be useful for some science cases (especially on bluer stars) and as a cross-check on the measurements from the red arm, and therefore we discuss more on blue arm spectra in Appendix \ref{sec:580v_rv_validation}.

All RVs reported in this paper are heliocentric velocities after the barycentric motion of the Sun is corrected, unless otherwise noted. 

\subsection{Radial Velocity Validation}
\label{sec:rv_validation}

The validation of RVs consists of comparing the radial velocities to external catalogues as well as assessing repeated observations within \SSSSS.

The cross-match of the \SSSSS dataset with the APOGEE DR14 data \response{contains $\sim$ 800 stars and} shows that the derived \SSSSS radial velocities have a systematic offset of 1.11\,\kms.\footnote{The cause of the offset is not yet clear and is likely related to either wavelength calibration bias, template mismatches or asymmetries in the line-spread function.} A similar offset is seen in the comparison with \gaia DR2 RVS velocities, therefore we subtract this offset and define our final RVs as $$
v_{S5} = v_{\mathrm{rvspecfit}} - 1.11\,\kms .
$$

As mentioned earlier, some stream fields were observed more than once if the first observation was taken in poor weather. Some stars were also observed repeatedly when the AAT fields overlapped.\footnote{This is because AAT has a FOV slightly larger than $2\degr$ in diameter.} We therefore are able to use those observations to assess the repeatability of RV measurements and the accuracy of RV uncertainties determined by the  pipeline.
Specifically, we consider all the pairs of repeated observations with RV uncertainties $\sigma_{v} < 30$\,\kms and S/N $>4$.  We then model the pair-wise radial velocity differences ${\delta v_{i,j}} = v_i - v_j$ by a Gaussian model combined with an outlier model $$ {\delta v_{i,j}} \sim f\, {\mathcal N}\left(0, \sqrt{F(\sigma_{v,i})^2 + F(\sigma_{v,j})^2}\right) + (1-f)\, {\mathcal N}(0,\sigma_{outl}) $$
where $\sigma_{v,i},\sigma_{v,j}$ are the RV uncertainties of the i-th and j-th observation respectively and 
$F(\sigma_v)=\sqrt{\sigma_{v,floor}^2 + (k\times \sigma_v)^2}$
is the uncertainty transformation function. Here $k$ is the scaling factor for the RV uncertainty and $\sigma_{v,floor}$ is the systematic floor of radial velocity precision. We fit the model to $\sim 3500 $ repeated observations and find $k=1.28$ and systematic floor is  $\sigma_{v,floor}=0.66$\,\kms. 
Thus our final RV uncertainties are determined as 
$$\sigma_{v,S5} = \sqrt{(1.28 \, \sigma_{v,\mathrm{rvspecfit}})^2 + 0.66^2}$$

We note that the likely reason for the presence of the systematic floor in RV determination is the accuracy of the 2dF/AAOmega wavelength calibration. The multiplicative constant in the radial velocity uncertainty is not equal to 1  probably because of the covariance between  pixels in the reduced spectra (produced naturally as a result of various rebinning/resampling steps of the 2dF pipeline). We find that the correlation coefficients of the noise between neighbouring pixels in the spectra are $\sim$ 0.3. If this covariance in the noise is ignored as it is in the current analysis, this is expected to produce  underestimated uncertainties by $\sim 30$\%, similar to what we empirically determine.

\begin{figure}
    \centering
    \includegraphics[width=0.5\textwidth]{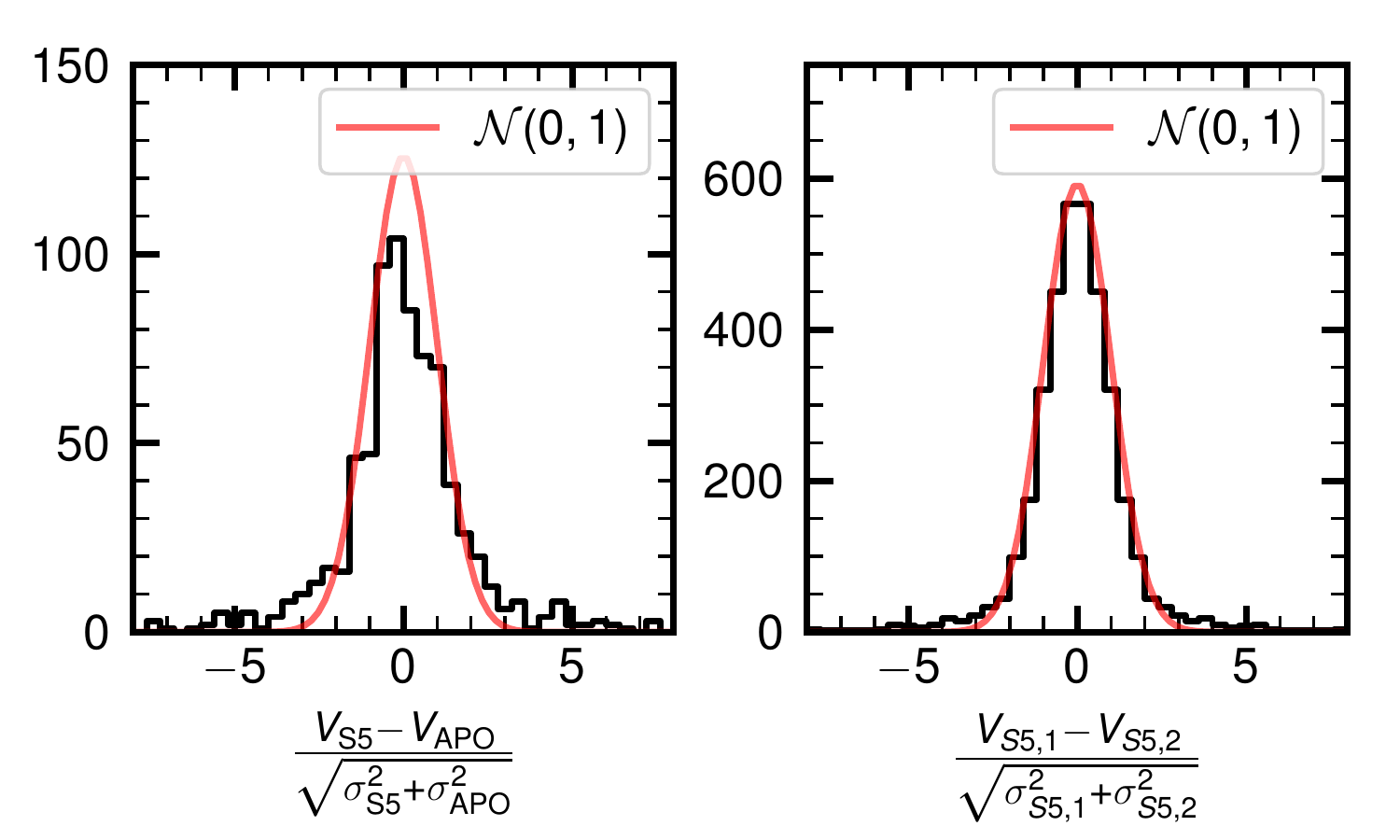}
    \caption{Comparison of \SSSSS radial velocities with APOGEE and repeated observations. {\it Left panel:} The black histogram shows the distribution of differences of the \SSSSS RVs and APOGEE RVs normalised by the combined uncertainty of \SSSSS and APOGEE ($\sqrt{\sigma_\mathrm{APO}^2 + \sigma_\mathrm{S5}^2}$). The red line shows the Gaussian distribution with zero mean unit variance. 
    The stars included for comparison have
    $(G_{BP}-G_{RP})_0<1.5$, small RV scatter in APOGEE $V_\mathrm{scatter}<0.5$\kms, $(\mathrm{S/N})_\mathrm{S5}>4$, and $\sigma_\mathrm{S5}<20\kms$. {\it Right:} The distribution of pairwise RV differences from \SSSSS repeated observations divided by the combined uncertainty $\sqrt{\sigma_\mathrm{S5,1}^2+ \sigma_\mathrm{S5,2}^2}$. The red curve shows the ${\mathcal N}(0,1)$ distribution that well describes the observations. We only show the stars with $\sigma_{v,\mathrm{S5}}<30$\kms and $\mathrm{S/N}>4$.}
    \label{fig:rvcompare}
\end{figure}

We demonstrate the performance of the recalibrated RVs and uncertainties in Figure~\ref{fig:rvcompare}. In the left panel we compare the \SSSSS RVs with APOGEE RVs (using the {\tt vhelio\_avg} column) by showing the distribution of \SSSSS and APOGEE RV differences normalized by the combined uncertainty $\sqrt{\sigma_{v, \rm S5}^2+\sigma_{v, \rm APO}^2}$. For the plot we use the APOGEE/\SSSSS stars that have high enough S/N>4, small RV errors ($\sigma_{\rm S5}<20$\kms in \SSSSS), and do not show significant RV variation in APOGEE ($v_\mathrm{scatter}<0.5$\,\kms). Since the APOGEE sample is dominated by red and cool objects, while the \SSSSS targets are significantly bluer on average, we additionally restrict our APOGEE/\SSSSS sample to stars with $(G_{BP}-G_{RP})_0<1.5$, which includes the majority  of stars (95\%) in \SSSSS. 
If the uncertainties of the \SSSSS RVs are correct and there are
no residual RV systematics,  the distribution shown on the left panel of Figure~\ref{fig:rvcompare} should behave like a ${\mathcal N}(0,1)$ Gaussian. The distribution is indeed centred at zero, with the core of the distribution similar to the ${\mathcal N}(0,1)$; however, more extended tails are also visible. The extended tails are likely caused by: (1) template mismatches and RV shifts related to convection or gravitational redshifts that can reach the level of $\sim$ 0.5<\kms \citep{Allende-Prieto:2013,Zwitter:2018}; (2) stellar binarity. While we remove stars that show RV variability in APOGEE $v_\mathrm{scatter}>0.5$\,\kms, it is likely that our sample contains longer period binaries with RV changes between the APOGEE and \SSSSS observations. 

The right panel of Figure~\ref{fig:rvcompare} assesses the repeatability of the radial velocities and correctness of the RV uncertainties by showing the distribution of pairwise RV differences in \SSSSS  divided by the combined uncertainty. Here the distribution is very close to the normal distribution with zero mean and unit variance, confirming the correctness of our error model and RV stability of our measurements.

In this section we described the validation of radial velocities determined from the red arm (1700D) that are used for the majority of the targets. We briefly discuss the same procedure for determining the zero-point offset and the error -model of the the blue arm (580V) RVs in Appendix~\ref{sec:580v_rv_validation}.

\subsection{[Fe/H] validation}
\label{sec:feh_validation}

To validate the \SSSSS \feh measurements we compare them with APOGEE, GALAH, GES, LAMOST and SEGUE survey data. We highlight that the \feh measurements are expected to be much more affected by systematic errors related to the stellar atmospheres/spectral templates used rather then purely random errors. Those systematic biases are also potentially different for stars with different atmospheric parameters, therefore we do not try to correct them but instead assess the overall quality of \feh measurements.  

First we adopt the same scaling for the \feh uncertainties as for the RVs, as it is caused by correlated noise in the spectra. 
$$\sigma_{\feh,S5} = 1.28 \times \sigma_{\mathrm{\feh,rvspecfit}}$$ 
This scaling also guarantees that repeated measurements  of \feh are consistent within the error (similar to right panel of Figure~\ref{fig:rvcompare}).

\begin{figure*}
    \centering
    \includegraphics[width=0.49\textwidth]{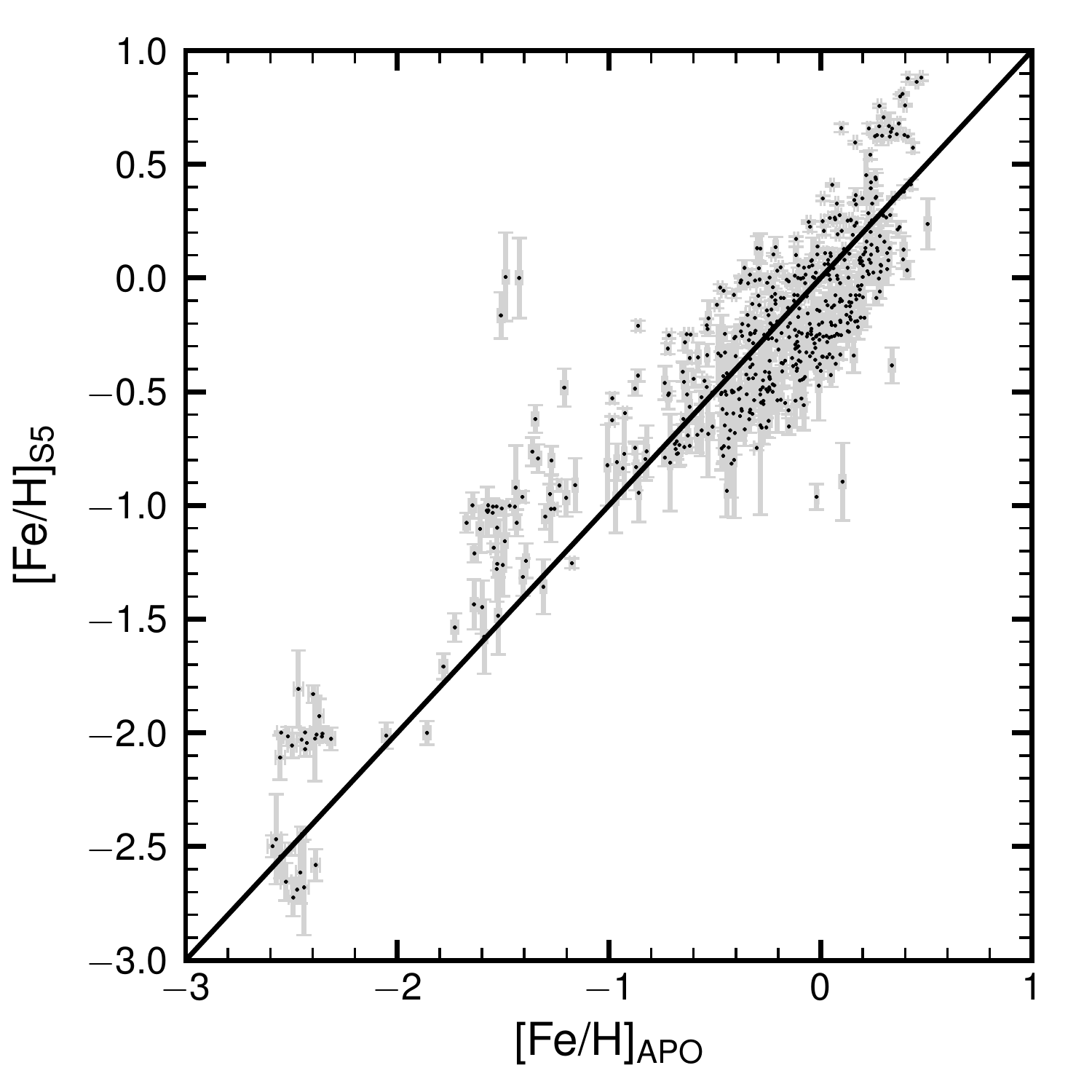}
    \includegraphics[width=0.49\textwidth]{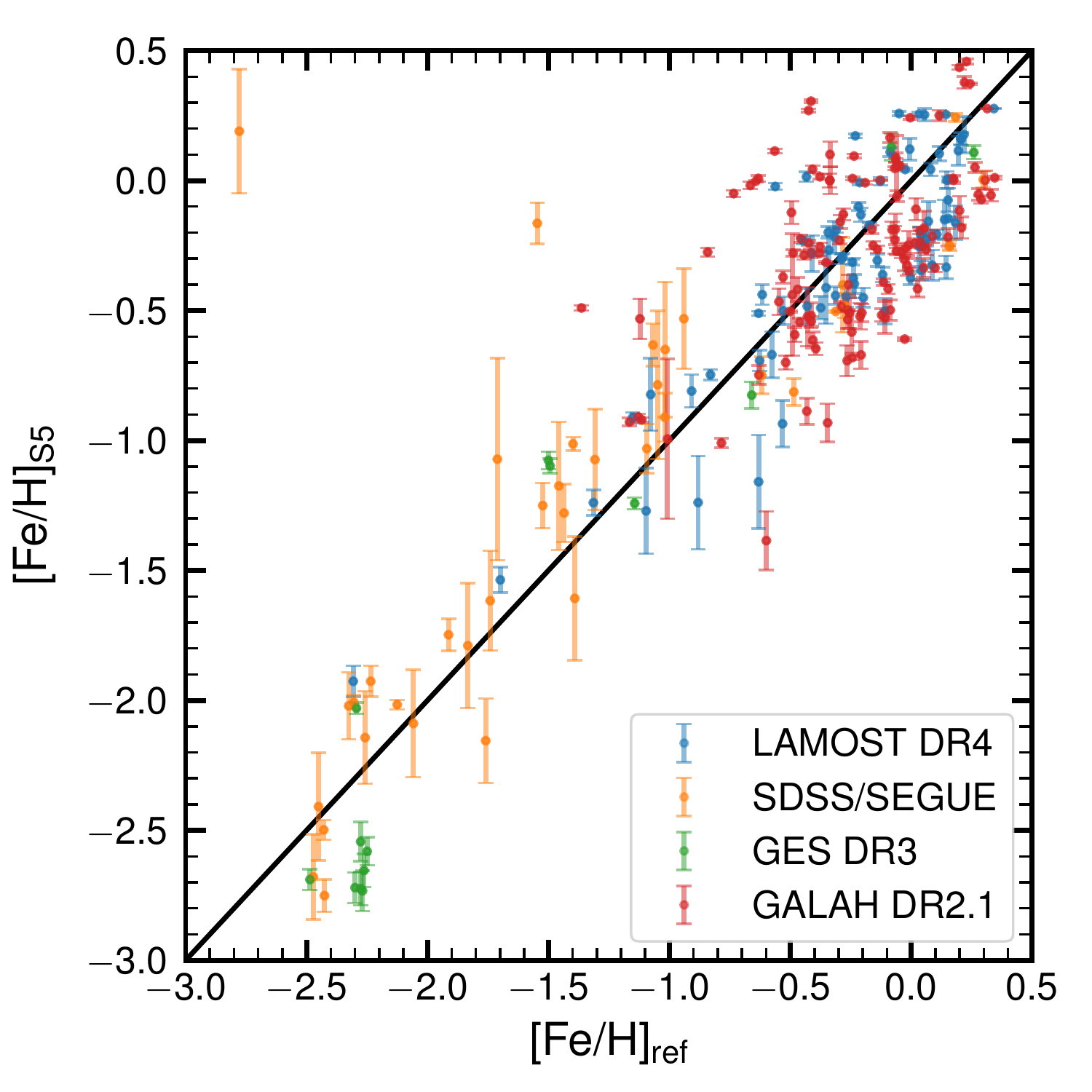}
    \caption{Comparison of the \SSSSS  spectroscopic metallicities from \code{rvspecfit} with APOGEE ({\it left}) and other surveys ({\it right}). We only use stars with $(G_{BP}-G_{RP})_0 <1.5$, S/N$>4$ and $\sigma_\mathrm{\feh,S5}<0.5$. The left panel additionally excludes the  APOGEE stars with effective temperatures below $\teff=4300$\,K. The right panel shows the comparison with various large spectroscopic survey datasets that were mostly serendipitously observed by \SSSSS. We notice that despite a few outliers and a spread of $\sim 0.3$\, dex there is a very good one-to-one mapping between our measurements and those from other surveys.
    %\JDS{The right panel needs different symbols as well as colours.}
    }
    \label{fig:fehcompare}
\end{figure*}

We start by looking at the comparison of the \feh from \code{rvscpecfit} (from the red arm spectra) with the APOGEE \feh.  We select the set of stars with both APOGEE and \SSSSS measurements similar to the one used in Section~\ref{sec:rv_validation}, but on top of that we also require that none of the {\tt STAR\_WARN} or {\tt STAR\_BAD} bits from APOGEE are set and that the $T_{\mathrm{eff,APO}}>4300$K, as we notice that for very cool stars (that are not representative of the \SSSSS targets) our pipeline produces a bias in the effective temperature and a bias in \feh. With this caveat in mind we compare with APOGEE abundances. The left panel of Figure~\ref{fig:fehcompare} shows the APOGEE metallicities vs \SSSSS metallicities. We remark that our metallicities track those from APOGEE, but with some occasional systematics; i.e. for very high metallicities  (\feh$>0.2$), there is a possible bias towards higher values (near \feh$\approx-1$). But overall the agreement is good with the systematic errors mostly below $\sim 0.2-0.3$\,dex and scatter of the same magnitude.

Since the APOGEE dataset is dominated by metal-rich giants we also compare the \SSSSS measurements with various large high and low resolution surveys  such as LAMOST, GALAH, GES and SEGUE. This is shown on the right panel of the Figure~\ref{fig:fehcompare}. Here we can see that with the additional surveys we get a much better sampling of the metal poor end of the stellar metallicity distribution, and we see no evidence of a significant metallicity bias. We note that there are some catastrophic outliers in the data as well.
Overall the summary of our metallicities with respect to various surveys as measured by the median deviation and half of the difference between 84\textsuperscript{th} and 16\textsuperscript{th} percentiles is $\{-0.18,0.34\}$ for GES, $\{0.10,0.33\}$ for GALAH, $\{-0.04,0.25\}$ for LAMOST, and $\{0.09,0.31\}$ for SDSS/SEGUE and $\{-0.02,0.21\}$ for APOGEE.

\subsection{Comparison with MIKE Spectroscopy at the metal-poor end}

Although we have shown the metallicity from \SSSSS is in good agreement with other surveys, the comparison set is largely metal-rich, while the stellar streams and stellar halo mostly consist of metal-poor stars. To verify the metallicity measurements from \SSSSS are robust on the most metal-poor stars in our sample, we observed a subset of the brightest stars ($g \lesssim 17$) with the high-resolution MIKE spectrograph \citep{Bernstein03} on the Magellan/Clay Telescope. MIKE targets were selected to be either stream members identified by \SSSSS or the extremely metal-poor star candidates from AAT metallicity of $\feh \lesssim -3.5$.

We observed our MIKE targets on 29-30 Sep 2018 with the 0\farcs7 slit in good weather, providing $R\sim30,000$ and $\sim40,000$ on the red and blue arms, respectively.
Data were reduced with the \code{CarPy} MIKE pipeline \citep{Kelson03}\footnote{\url{http://code.obs.carnegiescience.edu/carnegie-python-distribution}}.
Radial velocity measurement, continuum normalization, and equivalent widths of Fe\,I and Fe\,II lines were measured with a new version of the \textsc{smh} analysis environment first described in \citet{Casey14}\footnote{\url{https://github.com/andycasey/smhr}}.
A standard 1D LTE analysis was performed, using the ATLAS stellar atmospheres \citep{Castelli04} and the MOOG radiative transfer code updated to include scattering \citep{Sneden73,Sobeck11}\footnote{\url{https://github.com/alexji/moog17scat}}.
The effective temperature and microturbulence were determined by balancing the Fe\,I abundance vs. excitation potential and reduced equivalent width, respectively. The surface gravity was set by balancing the Fe\,I and Fe\,II abundances.
Following \citet{Frebel13}, we then corrected the effective temperature to match the photometric effective temperature scale (which for cool metal-poor giants typically increases \feh by ${\approx}0.2$ dex), and readjusted the surface gravity and microturbulence accordingly.
Statistical uncertainties were estimated from the error in the slopes for effective temperature and microturbulence, and combined standard error on the mean for surface gravity. We adopt the standard deviation of Fe\,I abundances as the \feh uncertainty.
Typical systematic uncertainties are 150\,K, 0.3 dex, 0.2 km\,s$^{-1}$, and 0.1 dex for effective temperature, surface gravity, microturbulence, and metallicity, respectively (see \citealt{Ji19} for details).

Here we are mostly interested in validating \SSSSS metallicities of the most metal-poor stars and therefore we only focus on the comparison of the metallicities from MIKE observations and from AAT observations. A full abundance analysis of other elements as well as the scientific interpretation of this data set will be presented in future work.

\begin{figure*}
    \centering
    \includegraphics[width=0.99\textwidth]{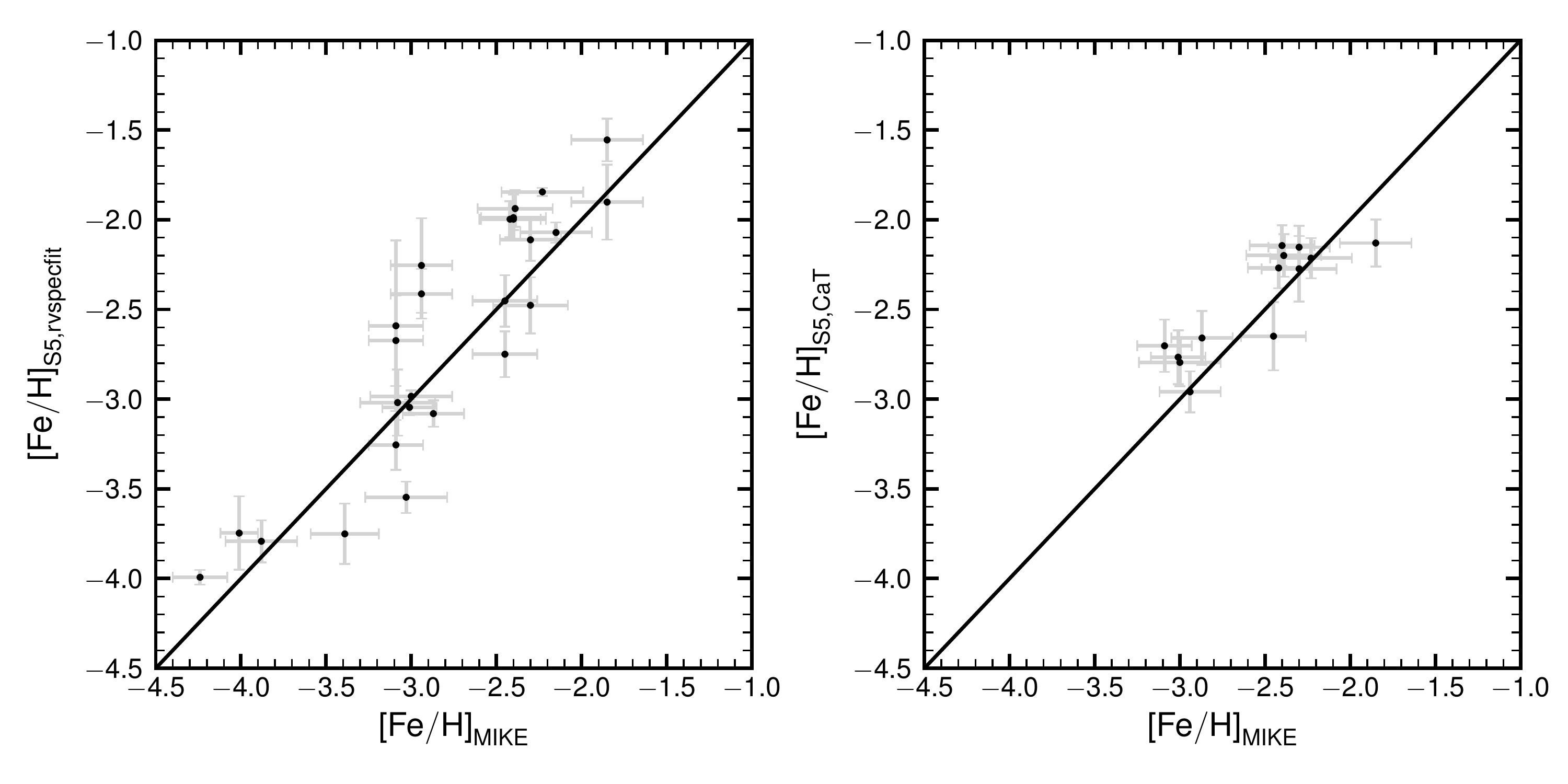}
    \caption{Comparison of the metallicities from AAT observations with MIKE spectroscopy for a subset of stream member stars and extreme metal-poor candidate stars. The left and right panels shows the AAT metallicities derived from \code{rvspecfit} and the CaT lines, respectively. Since the CaT metallicity requires distance as input, only stream members are shown. CaT metallicities in generally show a tighter sequence to the MIKE metallicities than the \code{rvspecfit} metallicity.}
    \label{fig:mikecompare}
\end{figure*}

Figure \ref{fig:mikecompare} shows the metallicity measurements from MIKE in comparison with the AAT observations. The left panel shows the metallicities derived from \code{rvspecfit} template fitting method with all MIKE targets. Despite a large metallicity range from $-4\lesssim\feh\lesssim-1.5$, the metallicities from the two independent measurements are in good agreement. In the right panel of the Figure, we compare with the CaT metallicities from AAT observations (Section \ref{sec:cat}). Since CaT metallicities require the distance of the star as an input, only stream members are shown. The $\feh_\mathrm{S5,CaT}$ vs. 
$\feh_\mathrm{MIKE}$ metallicity show a tighter sequence with an {\it rms} of 0.18 dex than the $\feh_\mathrm{S5,\code{rvspecfit}}$ vs. $\feh_\mathrm{MIKE}$ with an {\it rms} of 0.3 dex.

We therefore conclude that the metallicities derived from \code{rvspecfit} are generally reliable even at the most metal-poor end. However, if the distance of the star is known, the CaT metallicity exhibits smaller scatter. In future studies on stellar streams, CaT metallicities will be considered when available.

\subsection{QSOs in \SSSSS}
\label{sec:qsos}

During the visual inspection of galaxy redshifts as described in Section \ref{sec:galaxy-z}, we found that our stellar sample contains a large ($> 1000$) population of QSOs. 

\begin{figure}
    \centering
    \includegraphics[width=0.49\textwidth]{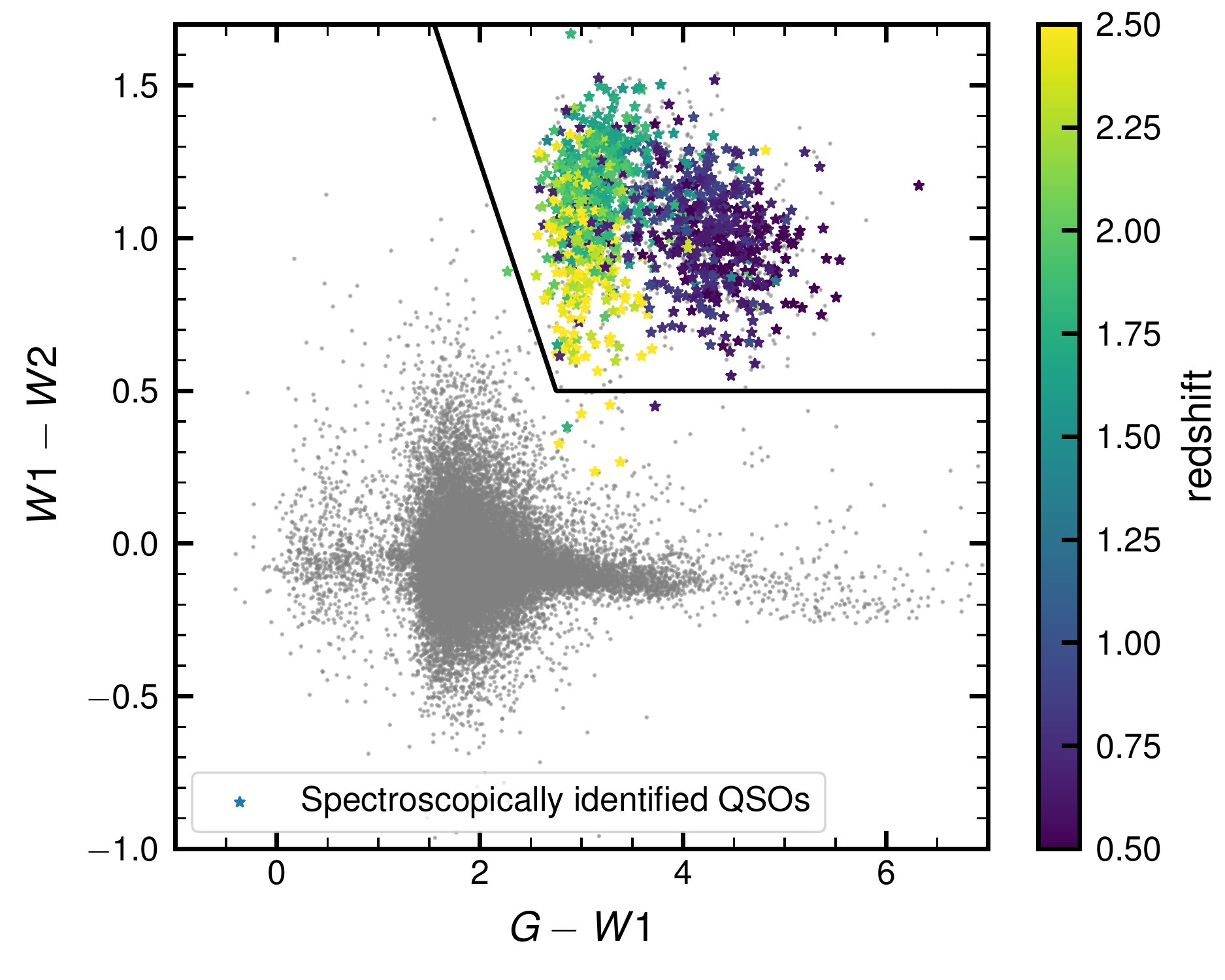}
    \caption{The \gaia-WISE colour--colour plot showing all the \SSSSS targets with coloured star symbols marking the subset of objects that were spectroscopically identified as QSOs. The colour indicates the  redshifts. The broken solid line denotes the boundary (Eq.~\ref{eq:qsosel}) that we adopt to reject the QSOs in the stellar analysis.}
    \label{fig:QSO_colors}
\end{figure}

To efficiently identify QSOs in the \SSSSS data, and remove them from stellar analyses, we use the combination of the WISE data with \gaia DR2 data, as they are known to be highly efficient for selecting QSOs \citep[see e.g.][]{Wright:2010, Lemon:2017}.
We crossmatch all the targets with the unWISE catalogue that combines the data from the original WISE mission and NEOWISE \citep{Schlafly:2019}. 
Figure~\ref{fig:QSO_colors} shows the Gaia-WISE colour $G-W1$ vs $W1-W2$ colour distribution of all \SSSSS targets, where the colour-coded symbols indicate objects that have been labeled as QSOs from visual inspection of the spectra and have measured redshifts, while grey dots show all other objects in \SSSSS. The colours of the QSO symbols correspond to the QSO redshift ($0.5 \lesssim z \lesssim 2.5$). It is clear that the QSOs occupy a well defined area of colour space, therefore for the stellar analysis we exclude all the objects lying above the broken line in the Figure defined as:
\begin{equation}
    (W1-W2) > {\rm Max} (0.5, 0.5 - (G-W1-2.75)) 
    \label{eq:qsosel}
\end{equation} 
which selects $\sim 1700$ QSO candidates in the current \SSSSS sample. We note that fewer than 10 out of $\sim1100$ spectroscopically identified QSOs lie outside of our QSO selection area.
For the spectroscopically identified QSOs with robust redshift measurement, we provide their redshifts in the table of Appendix \ref{sec:qsotable}.

\subsection{Selection of good quality stellar spectra}\label{sec:goodstar}

The \SSSSS data set spans a very large range of signal-to-noise, and includes contamination from galaxies, QSOs, and sometimes rare object types. We therefore need a way of identifying low quality, poorly fit stellar spectra or non-stellar spectra without individually examining each spectrum.
One way to do that is to define a threshold in the S/N, uncertainty in RV, and/or chi-squared values of the template fit. However, these are difficult criteria to define, since a high-metallicity cool star may have a good velocity determination even at very low S/N while a hot and/or metal-poor star may not get an RV measurement even at S/N=10. Furthermore, at high S/N the template mismatch to stellar spectra is very prominent (leading to high chi-squared values) while at low S/N the sky line residuals  can be very significant in the 1700D spectra. Therefore we train a random forest classifier \citep[RFC;][]{Breiman2001} to identify good-quality stellar spectra. We fit it separately to the red and blue arm spectra. The parameters that we use as features are the chi-squared values of the fit, radial velocity error, radial velocity posterior skewness and kurtosis, effective temperature, median signal to noise in the spectrum and relative median absolute deviation from the best fit template model Median(|Spec-Model|)/Median(Spec). The classifier is trained to identify good quality spectra, that were labeled using a |RV|$<$500\kms criterion, because many non-stellar or low quality and S/N objects are spuriously assigned to larger radial velocities up to the very edge of the considered range of 2000\kms. The RFC returns a probability of being a good stellar spectrum. This does not guarantee that all the stellar parameters are trustworthy, but provides a base selection of very likely stars with reliable RV. We then create a binary (0/1) flag \code{good\_star} and set it to 1 for  those with the RFC good spectrum probability $> 0.5$. Finally, we set \code{good\_star} = 0 for all objects that are identified as QSOs based on their WISE-\gaia photometry in Section \ref{sec:qsos} and galaxy targets in Section \ref{sec:galaxy-z} as non stellar objects.

The \code{good\_star} flagging removes both the non-stellar spectra and the stellar spectra with bad measurements (e.g., low S/N, etc.). We use the \code{good\_star} = 1 for the stellar-related science discussion in Section 5. We note, however, the selection might be chosen differently for future science paper upon the science goals.

\section{Early Science Results}\label{sec:science}

\subsection{Stream Results}

The primary goal of the \SSSSS survey is to confirm the newly discovered streams kinematically, as well as to measure the RVs and metallicities of the streams to understand their orbits, evolution, progenitors, etc. While we will provide more detailed studies of individual streams in future papers, here we give a brief overview of the stream kinematics based on the \SSSSS data collected so far. We focus on the DES streams since none of them were previously observed or confirmed spectroscopically. 

As shown in Figure~\ref{fig:s5pointing}, among all the DES streams that have been observed, Tucana III was observed entirely before \SSSSS and published in ~\citet{Li:2018b}; Turranbura has only one field observed so far; the other nine streams have been mostly covered in 2018. 

\begin{figure*}
    \centering
    \includegraphics[width=0.95\textwidth]{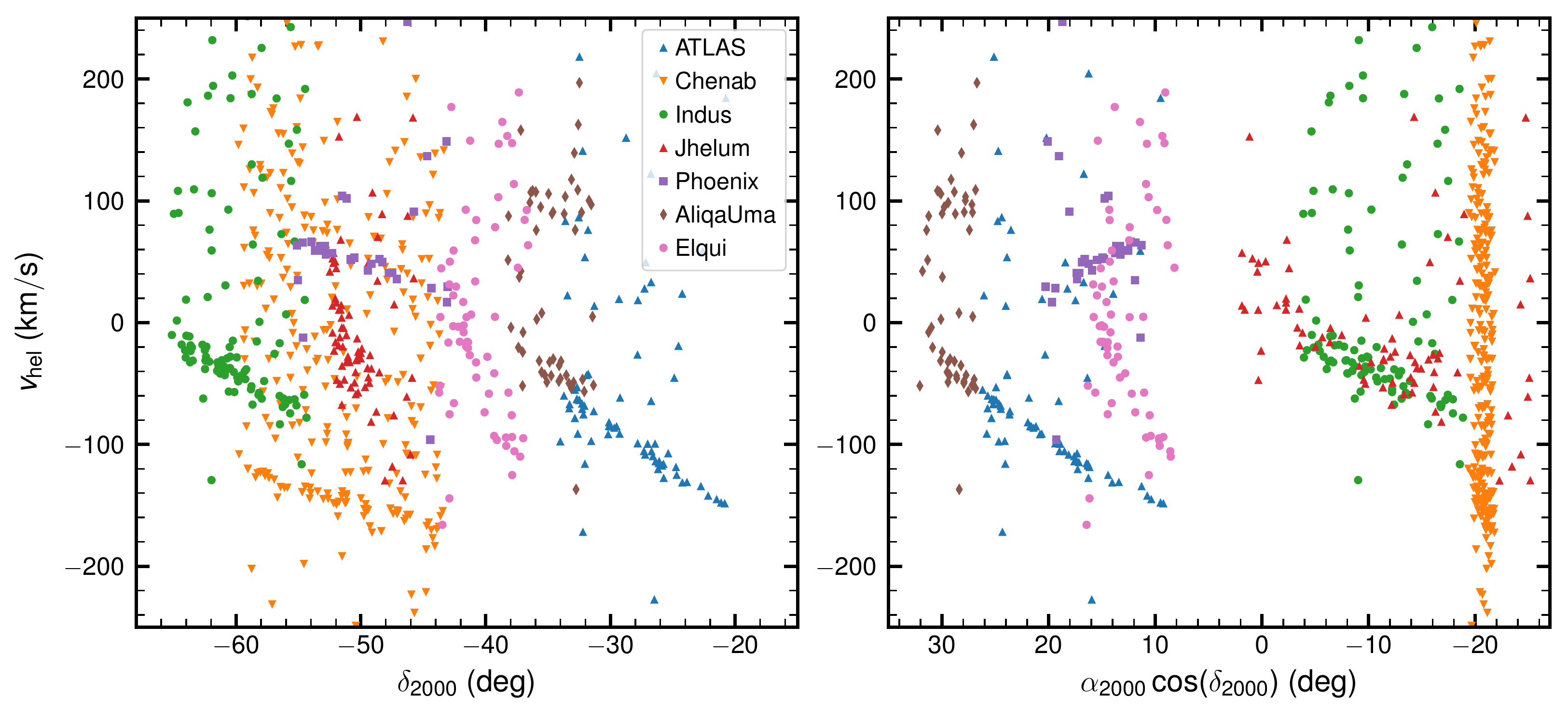}
    \caption{\response{Heliocentric RV as a function of declination ($\delta_{2000}$, {\emph left}) and right ascension corrected by declination ($\alpha_{2000}\cos{(\delta_{2000})}$, {\emph right})  for stream target stars}, colour coded for different stream fields. Seven streams with clear signals of stream members (i.e. clumpiness in RV) are shown.
    Plotting only high priority (P9) targets with $\logg < 4.1$ and $\feh < -1$ and no radial velocity cuts, the stream members are immediately visible. 
    }
    \label{fig:stream_rvs}
\end{figure*}

Figure~\ref{fig:stream_rvs} provides the summary of the stream kinematics as seen by \SSSSS. We plot the heliocentric RVs of the targeted stars as a function of declination for seven streams, for which we see a clear  kinematic signal (i.e. clustering in RV and a coherent change in RV as a function of declination). These seven streams are Aliqa Uma, ATLAS, Chenab\footnote{Here we treat Chenab and Orphan Stream separately although \citet{Koposov:2019} have shown that they are essentially one stream. We only show the stream targets in Chenab that are inside the DES footprint. A future paper will present all \SSSSS observations on Chenab+Orphan.}, Elqui, Indus, Jhelum, Phoenix. For the other two streams, Ravi and Willka Yaku, the member association does not immediately stand out, and further investigations or observations are needed.

Figure~\ref{fig:stream_rvs} highlights the abundance of cold substructure present in the Milky Way halo. The sample of stars plotted in Figure~\ref{fig:stream_rvs} has been selected using proper motion and photometric information and has no RV selection applied. RV measurements from \SSSSS produce a highly structured picture with individual streams clearly resolved. In future papers, we will investigate these streams in more details. In particular, two streams, the ATLAS stream and the Aliqa Uma stream (at $\delta_{2000}\sim -35\degr$ to $-20\degr$ in Figure~\ref{fig:stream_rvs}), which were previously thought to be independent, are in fact a single stream from phase space information (Li et al. in prep\@).

\subsection{BHB/BS Separation with $griz$ colour}\label{sec:bhb_bs_separation}

Old, metal-poor BHB stars have bright absolute magnitudes and are robust distance indicators, and thus are ideal tracers of the Milky Way's stellar halo~\citep{Deason:2011, Deason:2012}. In SDSS, multi-band photometry has been used to select BHB stars~\citep[see e.g.,][]{Yanny:2000, Deason:2011}, The selection relies on the $u-g$ colour, which provides a subtle distinction between BHB stars and blue straggler (BS) stars\footnote{We note that we call all of the hot dwarfs BSs here but in principle some of them could be hot young main-sequence stars as well.} at similar temperature but higher surface gravity. However, DES and many other ongoing imaging surveys (e.g. Hyper Supreme Cam, DECaLS, etc) do not include $u$-band photometry. Recent work has shown that a combination of $griz$ photometry alone can also differentiate BHB stars from BSs \citep[see e.g.,][]{Vickers:2012, Belokurov:2016, Deason:2018}. Here, we use the stellar parameters measured from \SSSSS to show clear BHB and BS sequences separated with DES DR1 photometry.

\begin{figure*}
    \centering
    \includegraphics[width=0.99\textwidth]{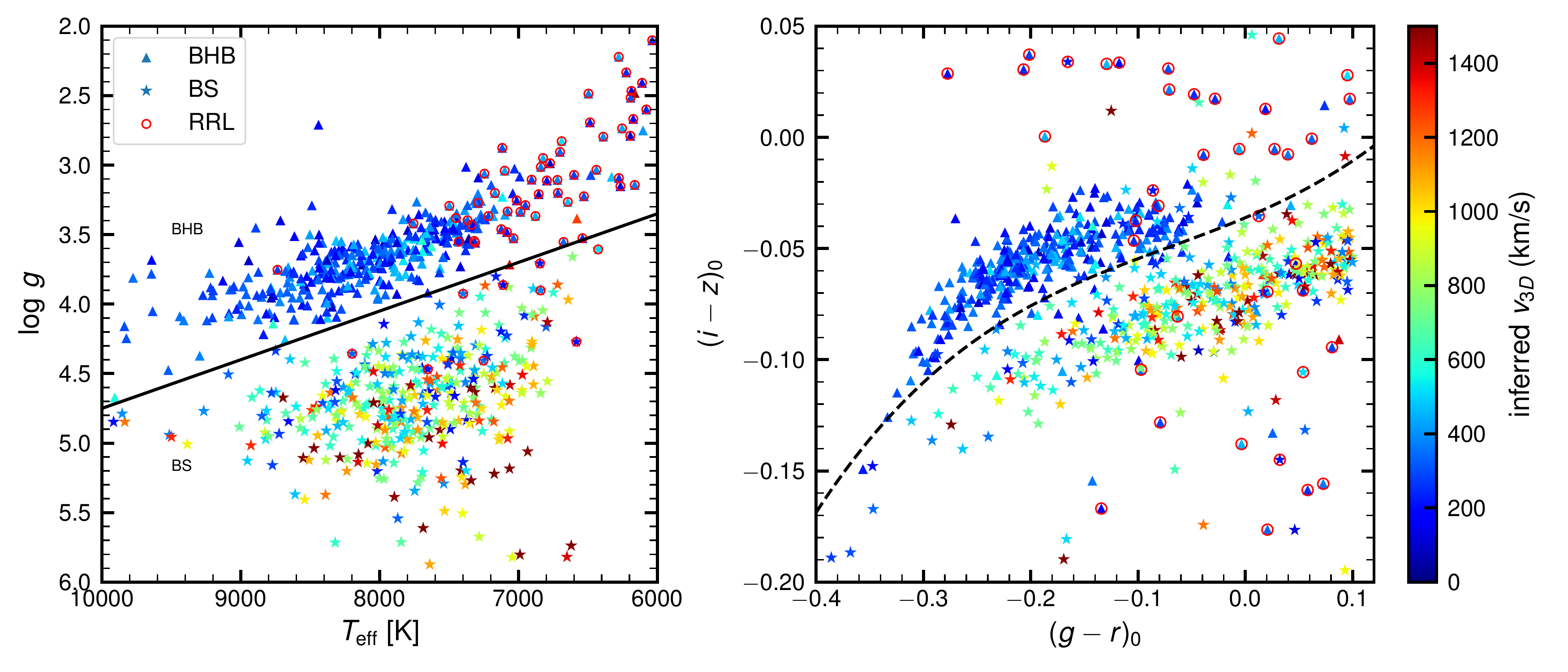}
    \caption{{\it Left panel}: The distribution of blue stars ($-0.4<(g-r)_0< 0.1$) in spectroscopic parameter \logg--\teff space. We can clearly see the separation between BHB and BS stars. We define our BHB (triangle symbols) and BS (star symbols) sample as stars below/above (respectively) the relation $\logg = 1.25 + 0.35\frac{\teff}{1000}$ shown by a black solid line. {\it Right panel:} The colour--colour distribution of \SSSSS blue stars. Thanks to the high photometric precision of DES DR1, BHBs and BSs are well separated in the dereddened $g-r$ vs $i-z$ diagram.  
    In both panels, all stars are colour-coded with the absolute value of their 3D heliocentric velocities, calculated with their distance based upon the assumption all of the stars are BHBs. 
    Therefore, their 3D velocities should be correct for the BHBs and inflated for the BSs. This effect is seen in the stellar parameter space (left panel) and colour-colour space (right panel).  
    The dashed curve on the right panel shows the polynomial in $(g-r)_0$ vs. $(i-z)_0$ that we propose to separate BHB from BS using photometry only (see Eq.~\ref{eq:bhb_bs_sep}).
    In addition, we identify the targeted RR Lyrae (RRL) stars from \gaia DR2 by open red circles in both panels. Although the BHBs and BS follow tight sequences in the colour--colour diagram, RRLs are scattered in colour depending on their phase at the time of observation, but are well clustered at $2\lesssim\logg\lesssim 4$ and $6000<\teff<7500$ in $\logg-\teff$ space.} 
    \label{fig:bhb_selection}
\end{figure*}

We select all stars in \SSSSS with $-0.4 <(g-r)_0 < 0.1$ (mostly from the P6 target class) and $6000 <\teff < 10000$~K as BHB candidates. In addition, we require \code{good\_star} = 1 and the S/N of the red arm spectra $>5$ to remove QSOs as well as to ensure the stellar parameters are reliable.  We then show in Figure \ref{fig:bhb_selection} that the BHBs and BSs are clearly distinct in (1) stellar parameters \logg--\teff space (left panel), (2) $(g-r)_0$ vs $(i-z)_0$ photometric space (right panel), and (3) kinematic space (colour coding). For the kinematics, we calculated the absolute value of the 3D heliocentric velocity $v_\mathrm{3D}$ of each star based on the proper motion from \gaia, RV from \SSSSS, and the heliocentric distance assuming that the star is a BHB.\footnote{The distance is derived using the BHB absolution magnitude relation $M_g$ vs $g-r$ in \citet{Belokurov:2016}.}
%Mg = 0.398 - 0.392*gr + 2.729*gr**2 + 29.1128*gr**3  + 113.569*gr**4 
Therefore, $v_\mathrm{3D}$ for BHBs is correct, while for most BSs that are intrinsically fainter and therefore closer, the inferred  $v_\mathrm{3D}$ is expected to be inflated by a factor of 2-3. This can be seen in the colour scale of the left panel of Figure showing that high \logg stars have high inferred $v_\mathrm{3D}$, further confirming that the stellar parameters from \SSSSS are robust to, e.g., separate the BHBs from BSs. We then select stars above the solid line
in the stellar parameter space (left panel) as BHBs (triangle symbols), and stars below  as BS (star-shaped symbols). These two populations show almost perfect separation in the colour--colour space in the right panel. 
The tight BHB sequence in $(g-r)_0$ vs $(i-z)_0$ allows high purity BHB selection with DES DR1 for any potential studies on the Milky Way halo. 

To make it is easier for future studies of BHBs we determine the curve that best separates the BHB from BS: 

\begin{eqnarray}
(i-z)_0 =& 
1.11371 (g-r)_0^5 -1.50963 (g-r)_0^4+ \nonumber\\
& 0.94966 (g-r)_0^3 +  0.29969 (g-r)_0^2+\nonumber\\
& 0.20021 (g-r)_0 -0.03684
\label{eq:bhb_bs_sep}
\end{eqnarray}
This curve is shown by a dashed line on Figure~\ref{fig:bhb_selection}, with BHB lying above, and BS below the curve. We have determined this function by fitting for the polynomial providing the maximum margin separation in $(i-z)$ between BHB and BS classes \citep[see e.g. section 7.1.1 of ][]{Bishop2006}. \response{As the separation between BHB and BS is also observed in the space of $g-r$ and $r-z$ colours, and surveys like DECaLS \citep{Dey:2018} do not observe in the $i-$band, we provide a boundary relying on $g$, $r$, $z$ filters alone:}

\begin{eqnarray}
(r-z)_0 =& 
1.07163 (g-r)_0^5 -1.42272 (g-r)_0^4+ \nonumber\\
& 0.69476 (g-r)_0^3 -0.12911 (g-r)_0^2+\nonumber\\
& 0.66993 (g-r)_0 -0.11368
\label{eq:bhb_bs_sep}
\end{eqnarray}

Both panels of Figure~\ref{fig:bhb_selection} also show the \gaia RR Lyrae (marked by red circles) that were targeted by \SSSSS. They are naturally scattered across the $(g-r)_0$ vs $(i-z)_0$ space due to variability, while mostly occupying a tight corner in stellar parameter space (low \logg and \teff$\lesssim 7500$~K).

We note that there are $\sim$ 400 BHB stars plotted in Figure \ref{fig:bhb_selection} (with an additional $\sim$400 BSs). In total, there are about 700 BHBs observed in \SSSSS (defined as $6500 < \teff < 10000$~K and above the solid line in the left panel of Figure \ref{fig:bhb_selection}). Among the $\sim300$ BHBs not shown in the Figure, about half are outside of DES footprint and therefore no $griz$ photometry is available, while the other half have S/N $< 5$ in the red arm spectra.
This BHB dataset is valuable for studies of the kinematic properties of the Milky Way halo, including measuring the potential of the Milky Way~\citep{Deason:2012}, studying the wake from LMC infall~\citep{Garavito-Camargo:2019,Belokurov:2019}, and understanding the remnants from \gaia Sausage in the outer halo~\citep[e.g.,][]{Lancaster:2019}. 

\subsection{Photometric Metallicity in $griz$ colour}\label{sec:mpdiscuss}

The broadband colours of stars are sensitive to the metallicity of the stars. The metallicity of a metal-poor star can usually be estimated via the ultraviolet (UV) excess, i.e. the difference between the star's $U-B$ colour and that which would be measured for a more metal-rich star with the same $B-V$ colour, \citep[see, e.g.,][]{Wildey1962, Sandage1969}.
\citet{Ivezic2008ApJ...684..287I} showed that the photometric metallicity of F/G stars could be estimated from the position of stars in the SDSS $u-g$ vs $g-r$ diagram. 
Unfortunately, DES does not routinely use the DECam $u$-band. However, \citet{Li:2018b} showed that metal-poor stars at $(g-r)_0 > 0.4$ are separable from metal-rich stars of the same $g-r$ colour using the $g-r$ vs $r-i$ colour combination instead. Here, we examine the correlation between the metallicity and broadband photometry in $griz$ with the stellar metallicities from \SSSSS data.

\begin{figure*}
    \centering
    \includegraphics[width=0.99\textwidth]{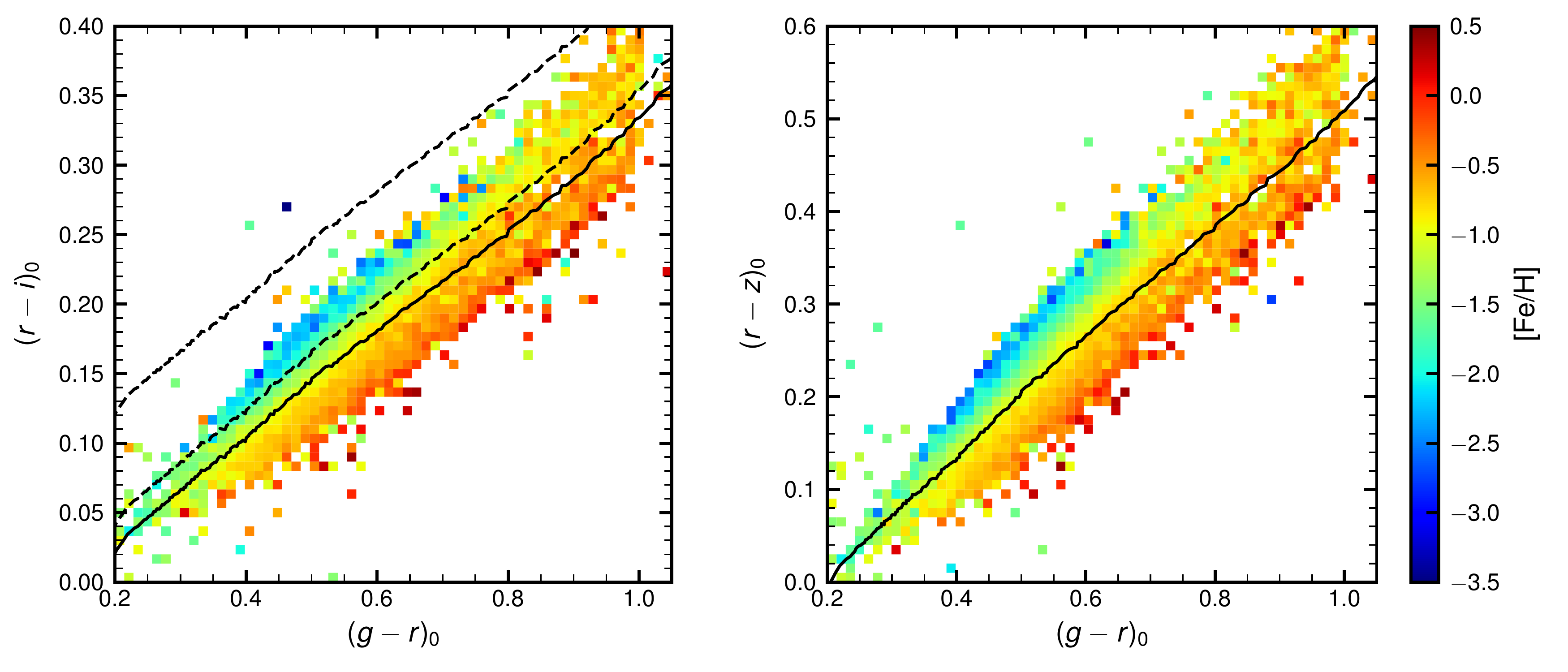}
    \caption{{\it Left panel}: 2D histogram in dereddened $g-r$ vs $r-i$ diagram, colour coded with the mean metallicity in each bin, obtained by \SSSSS. The solid line shows the stellar locus in DES. The dashed lines shows the boundary of the selection of metal-poor stars (P4 targets in Section \ref{sec:mp}). As discussed in the target selection for both streams and metal-poor stars, high precision DES DR1 photometry easily separates metal-rich stars from the metal-poor stars, with more metal-poor stars being located to the upper left of the stellar locus. Therefore, a photometric metallicity can be obtained based on $griz$ colour. Note, however, that such separation disappears at $(g-r)_0 \lesssim 0.4$ and therefore is not applicable to MSTO stars. {\it Right panel}: same plot but in dereddened $g-r$ vs $r-z$ space. Similar (and probably better) correlation between stellar metallicity and colours is visible.}
    \label{fig:feh_color}
\end{figure*}

We select stellar targets with \code{good\_star} = 1, S/N $ > 5$,  $\sigma_{\feh} < 0.5$ from \SSSSS to ensure that the sample has reliable metallicity measurements with small uncertainties. We limit this analysis to stars in DES footprint, which have $griz$ photometry from DES DR1. In Figure \ref{fig:feh_color}, we show the correlation of the stellar metallicity with the position of a star in the $(g-r)_0$ vs. $(r-i)_0$ diagram. The colour scheme shows the average metallicity value in each bin. At $0.4<(g-r)_0<0.8$, it is obvious that at a constant $(g-r)_0$ colour, metal-poor stars tends to have redder $(r-i)_0$ colours.
At $(g-r)_0 < 0.4$, the separation becomes weaker and therefore it is harder to assess the metallicity of MSTO stars with $gri$ colour alone.
A similar trend is also shown in the $(g-r)_0$ vs. $(r-z)_0$ diagram and therefore this correlation could be applied to DECaLS data, which lacks $i$-band observations, as well. 
In the future, photometric metallicities could be derived for the entire DES and DECaLS data sets, enabling statistical studies of the metallicity distribution of the Milky Way's stellar halo.

\subsection{Metallicity Distribution Function in \SSSSS}\label{sec:mdf}

\begin{figure*}
    \centering
    \includegraphics[width=0.9\textwidth]{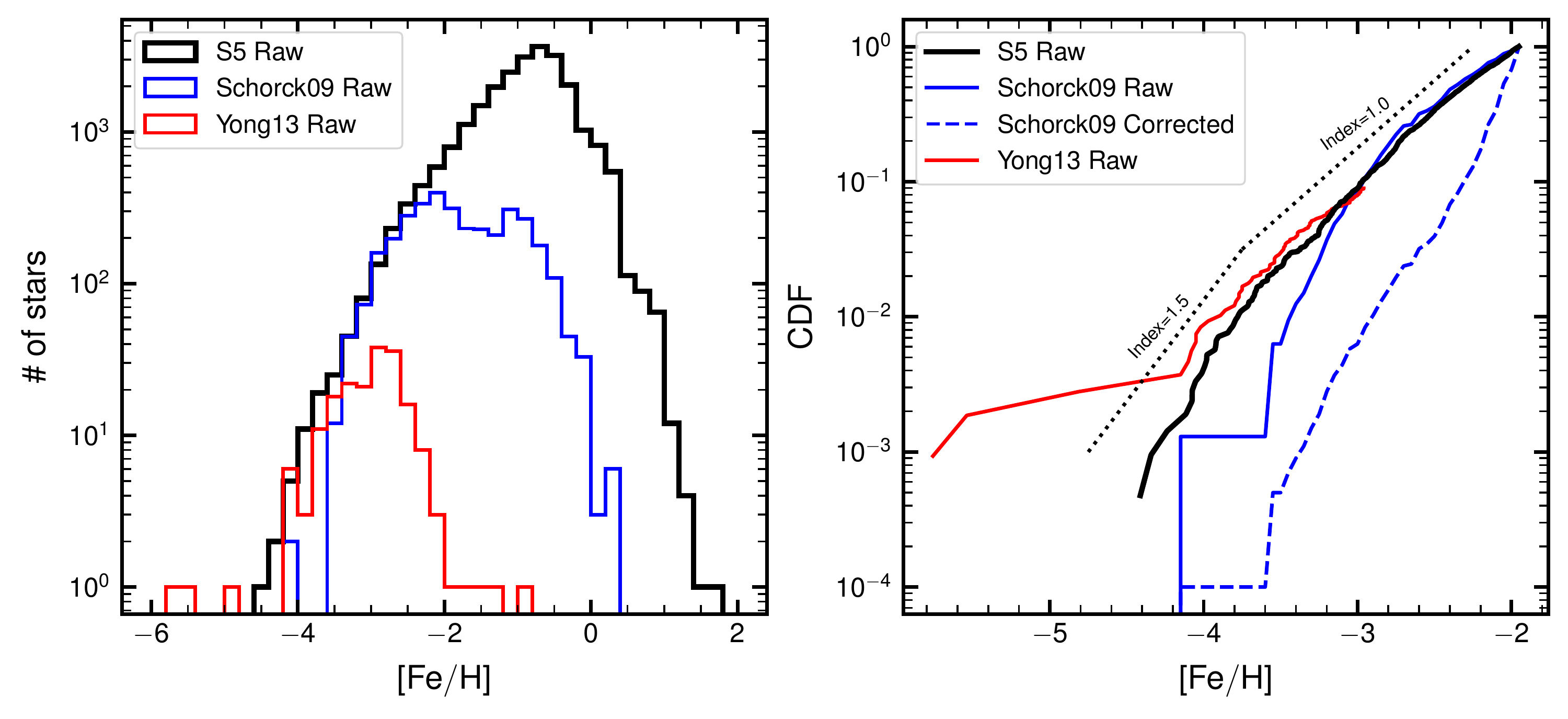}
    \caption{{\it Left panel}: Raw metallicity distribution function (MDF) from \SSSSS, compared to the HES \citep{Schorck2009} and a metal-poor star survey \citep{Yong2013}. \SSSSS has identified a comparable number of $\mbox{[Fe/H]} < -3$ stars and slightly more $\mbox{[Fe/H]} < -4$ than HES.
    {\it Right panel}: cumulative density functions (CDF) for \SSSSS, HES, and \citet{Yong2013}. The \citet{Yong2013} CDF only includes stars with $\mbox{[Fe/H]} < -3$, and has been rescaled to match \SSSSS at $\mbox{[Fe/H]} = -3$. Power law indices of 1.0 and 1.5 are shown for comparison. The dashed blue line shows the true MDF inferred by \citet{Schorck2009} after accounting for selection effects.}
    \label{fig:s5-mdf}
\end{figure*}

The \SSSSS target selection preferentially selected metal-poor halo stars, resulting in a large sample of stars with low metallicities.
%$\mbox{[Fe/H]} < -3$. 
In Figure~\ref{fig:s5-mdf}, we show the raw metallicity distribution function (MDF) from \SSSSS, after applying quality cuts of \code{good\_star} = 1, S/N $>5$ and $\sigma_\feh < 1$. 
In this sample, about 37\% (6\%) of the 22k stars have $\feh < -1$ ($\feh < -2$); $\sim$190 stars are at $\feh < -3$, of which $\sim$90 are brighter than $g_0 < 17.5$.
As an initial comparison, we show the raw MDFs from the Hamburg-ESO survey \citep[HES,][]{Schorck2009} and the metal-poor star survey by \citet{Yong2013}. The overall number of metal-poor stars in \SSSSS is comparable to these previous surveys, though the \SSSSS stars are fainter and may thus probe farther into the halo. These raw star counts should not be compared to models, since selection effects are extremely important (see dashed line in right panel of Figure~\ref{fig:s5-mdf}). Future work quantifying the \SSSSS selection function should provide interesting constraints on the Milky Way's MDF.

\section{Conclusions and Future Work}\label{sec:summary}

In this paper we present the Southern Stellar Stream Spectroscopic Survey (\SSSSS), targeting the kinematics and chemistry of stars in prominent tidal streams in the Galactic halo. The \SSSSS survey design combines observations of stellar streams, the primary objective of \SSSSS, with auxiliary observations of various Milky Way halo tracers and low-redshift galaxies. We show, by comparing our measurements with APOGEE and other spectroscopic surveys, that the \SSSSS radial velocities and stellar metallicities are robust.

As of June 2019, we have observed 110 AAT fields ($\sim 3$ deg$^2$ per field), of which 70 fields are in the DES footprint. We observed nine DES streams along $> 80$\% of their length. When combined with the previously published Tucana III stream, the \SSSSS data represent the first spectroscopic measurements of ten out of fourteen DES streams. Among the nine newly observed streams, we have confirmed at least seven of them as genuine streams in phase space, with each of them displaying small velocity dispersion (Figure \ref{fig:stream_rvs}). The observed velocity distribution of \SSSSS demonstrates that the Milky Way stellar halo is rich in substructure. Some of the structures that are not easily associated with each other based on positions on the sky line up in phase space, suggesting a common origin. This emphasizes the crucial role the upcoming spectroscopic surveys \citep[such as WEAVE, DESI, 4MOST, SDSS-V; ][]{Dalton:2012,DESI:2016,deJong:2012,Kollmeier2017}  will play in the mapping and untangling the Milky Way's stellar halo. 

So far, \SSSSS has collected 43k spectra on 38k unique targets, of which 3k are targeted as low-z galaxies, and the rest are targeted as stars.\footnote{However, 1.7k of the stellar targets turned out to be QSOs; see Section \ref{sec:qsos}.} Among 38k unique targets, 33k/25k/14k/5k of them have S/N $>$ 2/5/10/20 per pixel ($\sim0.23$~\AA/pixel) in the red arm spectra. 31k out of 38k spectra have \code{good\_star} = 1 (see definition in Section \ref{sec:goodstar}) and can be further used for stellar related science analysis.

In addition to the primary science goals related to stellar streams, \SSSSS has also collected spectra for a number of other targets that could potentially deliver interesting science results. For example,
\begin{itemize}
\item $\sim$400 RR Lyrae candidates were targeted, of which $\sim$340 have good spectra. 
\item $\sim$150 white dwarf candidates were observed.
\item $\sim$190 stars have measured $\feh < -3$, of which $\sim$90 have $g_0 < 17.5$.
\item $\sim$700 BHBs were identified (see Section \ref{sec:bhb_bs_separation}).
\item \response{One of the fastest hyper-velocity stars in the Galaxy (with a Galactocentric velocity of $> 1700$\kms) was discovered and associated with the ejection from the Galactic Centre \citep{Koposov2019hvs}.}
\item $\sim$1700 photometrically identified QSOs were observed, of which 640 have robust redshift measurements (see Section \ref{sec:qsos} and Appendix \ref{sec:qsotable} for details).
\item $\sim$3000 galaxy targets were observed, of which $\sim$2300 have robust redshift measurements and $\sim$300 are galaxies at $z < 0.05$ (the low-redshift ratio is $\sim$13\%, compared with 2\% expected for a magnitude-limited survey).

\end{itemize}

\SSSSS is an on-going survey. We have observed $\sim$25 nights in 2018 and 12 hrs in 2019 so far. More observations will be conducted in 2019. In addition to completing observations of the DES streams, we will also observe other streams in the Southern Hemisphere and Northern Hemisphere streams that are accessible from the Siding Spring Observatory. 

Several papers presenting the spectroscopic results on the stellar streams are in preparation, with additional papers on other auxiliary science topics, ranging from the discovery of a hyper velocity star to studies of the low-redshift galaxies.

This paper is based on the \SSSSS internal data release version 1.4. The first public data release of \SSSSS, containing all observations taken in 2018 and 2019, is scheduled for the end of 2020. Updates on \SSSSS can be found at \url{http://s5collab.github.io}.

\section*{Acknowledgements}

This paper includes data obtained with the Anglo-Australian Telescope in Australia. We acknowledge the traditional owners of the land on which the AAT stands, the Gamilaraay people, and pay our respects to elders past and present.
This paper includes data gathered with the 6.5 meter Magellan Telescopes located at Las Campanas Observatory, Chile. 

\response{The \SSSSS Collaboration gratefully acknowledges a time swap with the Pristine Collaboration that enabled the 12 hours \SSSSS observation in 2019. We also would like to thank Anke Arentsen for helpful discussions on flux loss in the 2dF fibres (see Appendix A). Finally, we would like to thank Nicholas Martin for reviewing this paper.}

% add grants here...
%TSL is supported by NASA through Hubble Fellowship grant HF2-51439.001 awarded by the Space Telescope Science Institute, which is operated by the Association of Universities for Research in Astronomy, Inc., for NASA, under contract NAS5-26555.
TSL, APJ, and YYM are supported by NASA through Hubble Fellowship grant HST-HF2-51439.001, HST-HF2-51393.001, and HST-HF2-51441.001 respectively, awarded by the Space Telescope Science Institute, which is operated by the Association of Universities for Research in Astronomy, Inc., for NASA, under contract NAS5-26555.
SK is partially supported by NSF grant AST-1813881 and Heising-Simons foundation grant 2018-1030.
J.~D.~Simpson, SLM and DBZ acknowledge the support of the Australian Research Council (ARC) through Discovery Project grant DP180101791.  
%APJ is supported by NASA through Hubble Fellowship grant HST-HF2-51393.001 awarded by the Space Telescope Science Institute, which is operated by the Association of Universities for Research in Astronomy, Inc., for NASA, under contract NAS5-26555.
YYM was supported by the Pittsburgh Particle Physics, Astrophysics and Cosmology Center through the Samuel P.\ Langley PITT PACC Postdoctoral Fellowship.
RHW, MG, and YYM received support from NSF AST-1517148.
G.~S.~Da~C. also acknowledges ARC support through Discovery Project grant DP150103294. 
J.~D.~Simon was supported in part by the National Science Foundation through grant AST-1714873. 

% SIMBAD
This research has made use of the SIMBAD database, operated at CDS, Strasbourg, France \citep{Simbad}, and NASA's Astrophysics Data System Bibliographic Services.

% Gaia
This work presents results from the European Space Agency (ESA) space
mission Gaia. Gaia data are being processed by the Gaia Data
Processing and Analysis Consortium (DPAC). Funding for the DPAC is
provided by national institutions, in particular the institutions
participating in the Gaia MultiLateral Agreement (MLA). The Gaia
mission website is https://www.cosmos.esa.int/gaia. The Gaia archive
website is https://archives.esac.esa.int/gaia.

%DES DR1
This project used public archival data from the Dark Energy Survey
(DES). Funding for the DES Projects has been provided by the
U.S. Department of Energy, the U.S. National Science Foundation, the
Ministry of Science and Education of Spain, the Science and Technology
Facilities Council of the United Kingdom, the Higher Education Funding
Council for England, the National Center for Supercomputing
Applications at the University of Illinois at Urbana-Champaign, the
Kavli Institute of Cosmological Physics at the University of Chicago,
the Center for Cosmology and Astro-Particle Physics at the Ohio State
University, the Mitchell Institute for Fundamental Physics and
Astronomy at Texas A\&M University, Financiadora de Estudos e
Projetos, Funda{\c c}{\~a}o Carlos Chagas Filho de Amparo {\`a}
Pesquisa do Estado do Rio de Janeiro, Conselho Nacional de
Desenvolvimento Cient{\'i}fico e Tecnol{\'o}gico and the
Minist{\'e}rio da Ci{\^e}ncia, Tecnologia e Inova{\c c}{\~a}o, the
Deutsche Forschungsgemeinschaft, and the Collaborating Institutions in
the Dark Energy Survey.  The Collaborating Institutions are Argonne
National Laboratory, the University of California at Santa Cruz, the
University of Cambridge, Centro de Investigaciones Energ{\'e}ticas,
Medioambientales y Tecnol{\'o}gicas-Madrid, the University of Chicago,
University College London, the DES-Brazil Consortium, the University
of Edinburgh, the Eidgen{\"o}ssische Technische Hochschule (ETH)
Z{\"u}rich, Fermi National Accelerator Laboratory, the University of
Illinois at Urbana-Champaign, the Institut de Ci{\`e}ncies de l'Espai
(IEEC/CSIC), the Institut de F{\'i}sica d'Altes Energies, Lawrence
Berkeley National Laboratory, the Ludwig-Maximilians Universit{\"a}t
M{\"u}nchen and the associated Excellence Cluster Universe, the
University of Michigan, the National Optical Astronomy Observatory,
the University of Nottingham, The Ohio State University, the OzDES
Membership Consortium, the University of Pennsylvania, the University
of Portsmouth, SLAC National Accelerator Laboratory, Stanford
University, the University of Sussex, and Texas A\&M University.
Based in part on observations at Cerro Tololo Inter-American
Observatory, National Optical Astronomy Observatory, which is operated
by the Association of Universities for Research in Astronomy (AURA)
under a cooperative agreement with the National Science Foundation.

%DECALS
The Legacy Surveys consist of three individual and complementary
projects: the Dark Energy Camera Legacy Survey (DECaLS; NOAO Proposal
ID \# 2014B-0404; PIs: David Schlegel and Arjun Dey), the
Beijing-Arizona Sky Survey (BASS; NOAO Proposal ID \# 2015A-0801; PIs:
Zhou Xu and Xiaohui Fan), and the Mayall z-band Legacy Survey (MzLS;
NOAO Proposal ID \# 2016A-0453; PI: Arjun Dey). DECaLS, BASS and MzLS
together include data obtained, respectively, at the Blanco telescope,
Cerro Tololo Inter-American Observatory, National Optical Astronomy
Observatory (NOAO); the Bok telescope, Steward Observatory, University
of Arizona; and the Mayall telescope, Kitt Peak National Observatory,
NOAO. The Legacy Surveys project is honored to be permitted to conduct
astronomical research on Iolkam Du'ag (Kitt Peak), a mountain with
particular significance to the Tohono O'odham Nation.

NOAO is operated by the Association of Universities for Research in
Astronomy (AURA) under a cooperative agreement with the National
Science Foundation.

The Legacy Survey team makes use of data products from the Near-Earth
Object Wide-field Infrared Survey Explorer (NEOWISE), which is a
project of the Jet Propulsion Laboratory/California Institute of
Technology. NEOWISE is funded by the National Aeronautics and Space
Administration.

The Legacy Surveys imaging of the DESI footprint is supported by the
Director, Office of Science, Office of High Energy Physics of the
U.S. Department of Energy under Contract No. DE-AC02-05CH1123, by the
National Energy Research Scientific Computing Center, a DOE Office of
Science User Facility under the same contract; and by the
U.S. National Science Foundation, Division of Astronomical Sciences
under Contract No. AST-0950945 to NOAO.

The national facility capability for SkyMapper has been funded through ARC LIEF grant LE130100104 from the Australian Research Council, awarded to the University of Sydney, the Australian National University, Swinburne University of Technology, the University of Queensland, the University of Western Australia, the University of Melbourne, Curtin University of Technology, Monash University and the Australian Astronomical Observatory. SkyMapper is owned and operated by The Australian National University's Research School of Astronomy and Astrophysics. The survey data were processed and provided by the SkyMapper Team at ANU. The SkyMapper node of the All-Sky Virtual Observatory (ASVO) is hosted at the National Computational Infrastructure (NCI). Development and support the SkyMapper node of the ASVO has been funded in part by Astronomy Australia Limited (AAL) and the Australian Government through the Commonwealth's Education Investment Fund (EIF) and National Collaborative Research Infrastructure Strategy (NCRIS), particularly the National eResearch Collaboration Tools and Resources (NeCTAR) and the Australian National Data Service Projects (ANDS).

%Pan-STARRS
The Pan-STARRS1 Surveys (PS1) and the PS1 public science archive have been made possible through contributions by the Institute for Astronomy, the University of Hawaii, the Pan-STARRS Project Office, the Max-Planck Society and its participating institutes, the Max Planck Institute for Astronomy, Heidelberg and the Max Planck Institute for Extraterrestrial Physics, Garching, The Johns Hopkins University, Durham University, the University of Edinburgh, the Queen's University Belfast, the Harvard-Smithsonian Center for Astrophysics, the Las Cumbres Observatory Global Telescope Network Incorporated, the National Central University of Taiwan, the Space Telescope Science Institute, the National Aeronautics and Space Administration under Grant No. NNX08AR22G issued through the Planetary Science Division of the NASA Science Mission Directorate, the National Science Foundation Grant No. AST-1238877, the University of Maryland, Eotvos Lorand University (ELTE), the Los Alamos National Laboratory, and the Gordon and Betty Moore Foundation.

% White dwarf selection made use of SDSS-III data...
Funding for SDSS-III has been provided by the Alfred P. Sloan Foundation, the Participating Institutions, the National Science Foundation, and the U.S. Department of Energy Office of Science. The SDSS-III web site is http://www.sdss3.org/.

SDSS-III is managed by the Astrophysical Research Consortium for the Participating Institutions of the SDSS-III Collaboration including the University of Arizona, the Brazilian Participation Group, Brookhaven National Laboratory, Carnegie Mellon University, University of Florida, the French Participation Group, the German Participation Group, Harvard University, the Instituto de Astrofisica de Canarias, the Michigan State/Notre Dame/JINA Participation Group, Johns Hopkins University, Lawrence Berkeley National Laboratory, Max Planck Institute for Astrophysics, Max Planck Institute for Extraterrestrial Physics, New Mexico State University, New York University, Ohio State University, Pennsylvania State University, University of Portsmouth, Princeton University, the Spanish Participation Group, University of Tokyo, University of Utah, Vanderbilt University, University of Virginia, University of Washington, and Yale University. 

This manuscript has been authored by Fermi Research Alliance, LLC under Contract No. DE-AC02-07CH11359 with the U.S. Department of Energy, Office of Science, Office of High Energy Physics. The United States Government retains and the publisher, by accepting the article for publication, acknowledges that the United States Government retains a non-exclusive, paid-up, irrevocable, world-wide license to publish or reproduce the published form of this manuscript, or allow others to do so, for United States Government purposes.

{\it Facilities:} 
{Anglo-Australian Telescope (AAOmega+2dF); Magellan/Clay (MIKE)}

{\it Software:} 
{\code{numpy} \citep{numpy}, \code{scipy} \citep{scipy}, \code{matplotlib} \citep{matplotlib}, \code{astropy} \citep{astropy,astropy:2018}, \code{emcee} \citep{Foreman_Mackey:2013},  \code{CarPy} \citep{Kelson03}, \code{moog} \citep{Sneden73,Sobeck11}, \code{q3c} \citep{Koposov2006}, \code{sklearn} \citep{Pedregosa2011}, \code{RVSpecFit} \citep{rvspecfit}}

%%%%%%%%%%%%%%%%%%%%%%%%%%%%%%%%%%%%%%%%%%%%%%%%%%



\bibliographystyle{mnras}
\bibliography{main}

%\vspace{0.4cm}
%\\
%\parbox{\textwidth}{
\section*{Affiliations}
{\small\it
\noindent
%\scriptsize
$^{1}$ Observatories of the Carnegie Institution for Science, 813 Santa Barbara St., Pasadena, CA 91101, USA\\
$^{2}$ Department of Astrophysical Sciences, Princeton University, Princeton, NJ 08544, USA\\
$^{3}$ Fermi National Accelerator Laboratory, P.O.\ Box 500, Batavia, IL 60510, USA\\
$^{4}$ Kavli Institute for Cosmological Physics, University of Chicago, Chicago, IL 60637, USA\\
$^{5}$ McWilliams Center for Cosmology, Carnegie Mellon University, 5000 Forbes Ave, Pittsburgh, PA 15213, USA\\
$^{6}$ Institute of Astronomy, University of Cambridge, Madingley Road, Cambridge CB3 0HA, UK\\
$^{7}$ Department of Physics \& Astronomy, Macquarie University, Sydney, NSW 2109, Australia\\
$^{8}$ Macquarie University Research Centre for Astronomy, Astrophysics \& Astrophotonics, Sydney, NSW 2109, Australia\\
$^{9}$ Sydney Institute for Astronomy, School of Physics, A28, The University of Sydney, NSW 2006, Australia\\
$^{10}$ Lowell Observatory, 1400 W Mars Hill Rd, Flagstaff,  AZ 86001, USA\\
$^{11}$ Australian Astronomical Optics, Faculty of Science and Engineering, Macquarie University, Macquarie Park, NSW 2113, Australia\\
$^{12}$ School of Physics, UNSW, Sydney, NSW 2052, Australia\\
$^{13}$ Department of Astronomy \& Astrophysics, University of Chicago, 5640 S Ellis Avenue, Chicago, IL 60637, USA\\
$^{14}$ Department of Physics and Astronomy, Rutgers, The State University of New Jersey, Piscataway, NJ 08854, USA\\
$^{15}$ Department of Physics and Astronomy, University of Pittsburgh, Pittsburgh, PA 15260, USA\\
$^{16}$ Pittsburgh Particle Physics, Astrophysics, and Cosmology Center (PITT PACC), University of Pittsburgh, Pittsburgh, PA 15260, USA\\
$^{17}$ Department of Astronomy, Yale University, New Haven, CT 06520, USA\\
$^{18}$ George P. and Cynthia Woods Mitchell Institute for Fundamental Physics and Astronomy, and Department of Physics and Astronomy, Texas A\&M University, College Station, TX 77843, USA\\
$^{19}$ Research School of Astronomy and Astrophysics, Australian National University, Canberra, ACT 2611, Australia\\
$^{20}$ Department of Physics, University of Surrey, Guildford GU2 7XH, UK\\
$^{21}$ Centre for Astrophysics and Supercomputing, Swinburne University of Technology		PO box 218 Hawthorn Vic 3122 Australia\\
$^{22}$ Centre of Excellence for All-Sky Astrophysics in Three Dimensions (ASTRO 3D), Australia\\
$^{23}$ Department of Physics, University of Wisconsin-Madison, 1150 University Avenue Madison, WI 53706, USA\\
$^{24}$ Kapteyn Instituut, Rijksunversiteit Groningen, the Netherlands\\
$^{25}$ School of Physics and Astronomy, Monash University, Wellington Rd, Clayton 3800, Victoria, Australia\\
$^{26}$ Cerro Tololo Inter-American Observatory, National Optical Astronomy Observatory, Casilla 603, La Serena, Chile\\
$^{27}$ Department of Physics, Stanford University, 382 Via Pueblo Mall, Stanford, CA 94305, USA\\
$^{28}$ Kavli Institute for Particle Astrophysics \& Cosmology, P.O.\ Box 2450, Stanford University, Stanford, CA 94305, USA\\
$^{29}$ SLAC National Accelerator Laboratory, Menlo Park, CA 94025, USA\\
}

%\begin{thebibliography}{99}
%\end{thebibliography}

%%%%%%%%%%%%%%%%%%%%%%%%%%%%%%%%%%%%%%%%%%%%%%%%%%

%%%%%%%%%%%%%%%%% APPENDICES %%%%%%%%%%%%%%%%%%%%%
\clearpage

\appendix

\section{Fibre positioning accuracy}\label{app:pointing} 

\begin{figure}
\centerline{\includegraphics[width=3.5in]{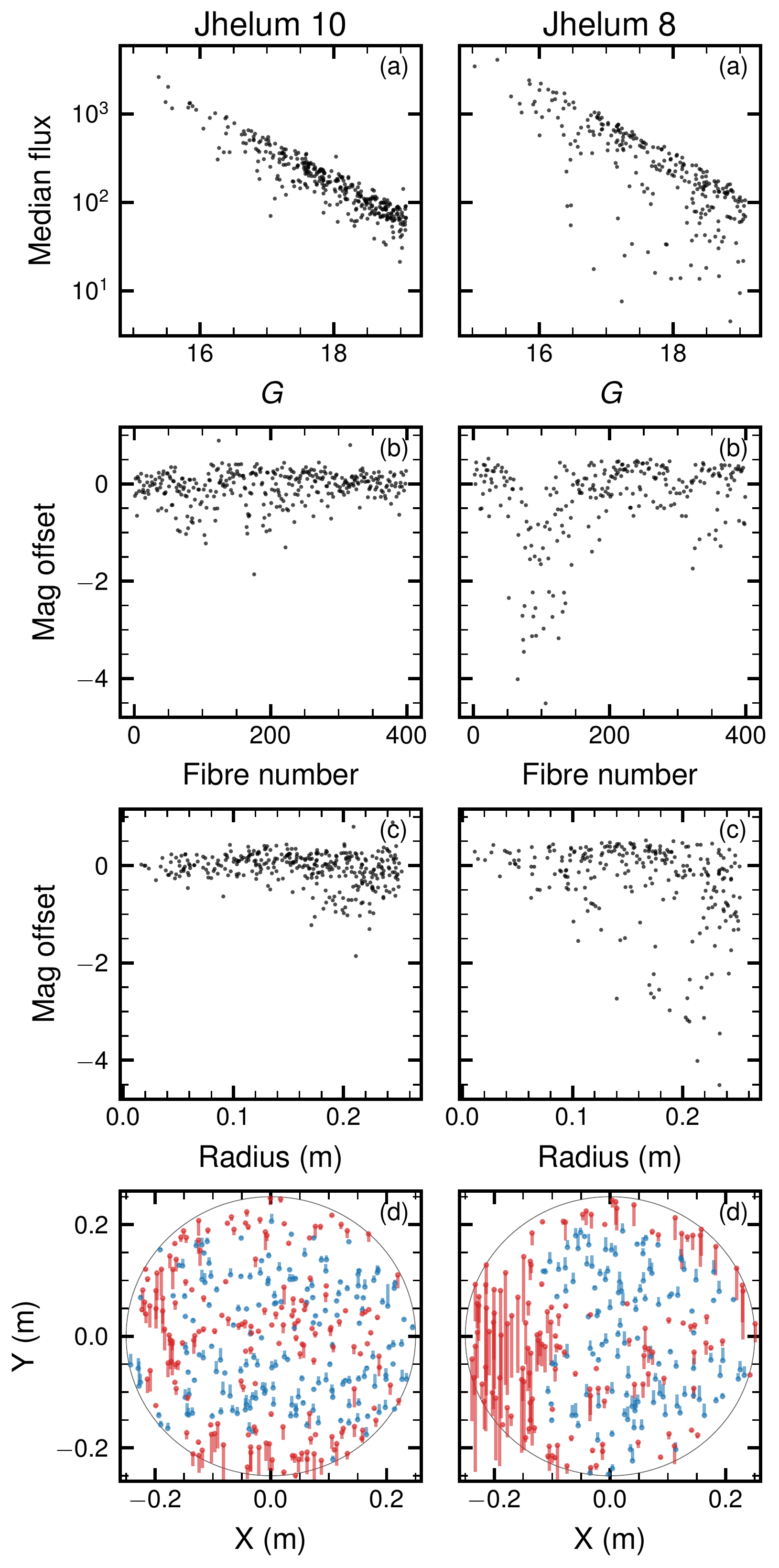}}
\caption{Comparison of the median spectral flux of the stream targets in Jhelum 10 (left column) and Jhelum 8 (right column) to their \gaia $G$ magnitude. In (a) we show the trend of magnitude with median flux, where for Jhelum 8 a number of stars have much lower flux than would be expected. (b) and (c) gives the difference in magnitudes between the $G$ magnitude and the median flux converted to a magnitude with respect to their fibre number and radius on the plate respectively. Stars above zero have more signal than expected, stars below have less signal than expected.  In (d) we show the X-Y position on the plate with a dot, and then a blue line up or a red line down to indicate if the star had more or less flux than expected. For Jhelum 10, there is little structure of trends with position on the plate or fibre used. But for Jhelum 8, there is a clear region of the plate with apparently poor positioning. 
}
\label{fig:quality_plots}
\end{figure}

The Two-degree Field (2dF) fibre positioner has been a workhorse instrument of the Anglo-Australian Telescope (AAT) for over 20~years. In its current iteration, a robotic X-Y gantry picks and places magnetic buttons onto the metal field plate, with these buttons attached to about 40 metres of optical fibres. These fibres transmits the light from the focal plane of the telescope to either the AAOmega or HERMES spectrographs. 

As with any fibre system, there are fibre-to-fibre throughput variations with 2dF \citep{Sharp:2013cs, Simpson:2016fk}. As discussed in \citet{Sharp:2013cs} light loss can occur at many places along the light path: at the primary mirror, at the air-fibre interface, and within the fibre itself. \citet{Simpson:2016fk} found that HERMES appeared to have fibre-to-fibre throughput variations at the fibre-spectrograph interface.

An additional source of lower-than-expected throughput is the fibre placement accuracy. If the fibre is placed in the focal plane off the position of the target, then less flux will be received at the spectrograph. The fibres observe about 2~arcsec of the sky (similar to the typical seeing at the AAT) and 2dF has a stated ability to place fibres to with an accuracy of $0.3$~arcsec on the sky \citep{Green2017}. However, the transformation from celestial coordinates to X-Y Cartesian positions for the robot gantry is complicated and requires careful calibration \citep{Cannon2008}.

We found that in some of our fields that regions of the plate appeared to have lower than expected signal. Here we compare the median flux of a given spectrum to the \gaia G magnitude of the star. We limit this comparison to only the expected stream candidate stars. This is because we wish to exclude the blue stars (which would have a different spectral energy distribution (SED) in the infrared than red stars), and galaxies (for which the magnitudes were taken from a different photometric system; and they are also not our main targets of interest). For a given star, the median flux in the reduced spectrum was converted to a magnitude\footnote{We used $-2.5\log(F)$, which does exclude stars with negative flux.} and subtracted from the \gaia magnitude of the star. A zero-point correction was applied equal to the median of all the magnitude difference in a given field. This analysis assumes that the SEDs of all our stars are similar enough that their flux in the near-IR can be compared to their $G$ magnitude. Some of the fibre-to-fibre scatter will likely be caused by the colour distribution of the stars.

In the left column of Figure \ref{fig:quality_plots} we show the results for the Jheluum 10 field. This can be considered a well-behaved field. The median flux follow the trend with stellar magnitude, and there are no obvious effects on throughput with fibre number. There is a slight decrease in throughput at larger radius on the plate, but this is to be expected as the plate scale decreases at the edge. In the bottom panel we show the X-Y position on the plate of the star with a dot, and then a blue line up or a red line down to indicate if the star had more or less flux than expected. The length of this line is scaled to how large the magnitude offset was. There is a small amount of clumpiness of the distributions of stars with higher or lower than expected flux in the X-Y plane.

In the right column of Figure \ref{fig:quality_plots} is a badly behaved field, Jhelum 8. Some of the stars have much lower flux than would be expected for their magnitude. These are concentrated with fibre numbers around 75--125 and are located at the edge of the plate. In the bottom panel of the column, these stars are all located in the left-hand edge of the plate. Because of the highly spatial nature of this on the plate, it is very likely to be an issue with the fibre placement, rather than other causes of fibre-to-fibre throughput variation.

We highlight this issue as it could potentially cause incompleteness in detecting the stream members. 
%We have assumed that stars at the faint end have all been observed to the same signal-to-noise ratio (modulo seeing, weather, and moon phase).
We emphasize that the $G-$band magnitude at S/N = 5 listed in Table \ref{table:fields} is the average magnitude over one AAT field. Due to the problem described above, we expect some fields have a number of stars brighter than the listed magnitude but with low S/N.
As such we caution against drawing any assumptions related to the spatial clumpiness or gaps of our streams in terms of our detected members.

\section{B. QSO Redshifts}\label{sec:qsotable} 

Over 1000 QSOs were identified spectroscopically via visual inspection. As QSO detection is not one of the main science goals of \SSSSS, we list in Table \ref{table:qso} 674 QSOs that have secure redshifts. We note that an additional 412 QSOs were spectroscopically identified, but the limited spectral coverage of \SSSSS included only a single broad emission line, and therefore the redshifts could not be unambiguously determined. The spectra of all QSOs are available upon request.

\begin{table}
\begin{center}
\caption{
A list of 674 QSOs identified by \SSSSS with robust redshift measurements.
Columns from left to right are \gaia DR2 Source ID, RA, Declination, $G$-band magnitude, and measured redshift.
The full table is available in the online version in machine readable format.
}
\scriptsize
\label{table:qso}
\begin{tabular}{l r r r r}
\hline
\gaia DR2 Source ID  & RA (deg)  & Decl. (deg) & $G$ (mag) & redshift \\
\hline
\hline
 4972649539728306432 &   4.697280  & -51.759919  & 18.5  & 0.44   \\
 2362400547317486720 &   9.358120  & -20.343325  & 19.1  & 1.85   \\
 2350270460161251200 &   9.375939  & -21.322281  & 19.8  & 2.48   \\
 2350251115628542464 &   9.435354  & -21.525050  & 19.9  & 1.82   \\
 2362705077678071424 &   9.917326  & -20.550277  & 20.0  & 2.57   \\
 5001051196383860352 &  10.270612  & -37.155282  & 19.6  & 0.72   \\
 5000985083952286336 &  10.317996  & -37.613705  & 19.1  & 0.88   \\
 5001096894835994880 &  10.349795  & -36.705281  & 19.4  & 2.24   \\
 2349979811134793088 &  10.383027  & -21.975567  & 18.6  & 2.72   \\
 2349914531927728640 &  10.478138  & -22.344380  & 19.5  & 3.31   \\
 2350032690772566016 &  10.534402  & -21.499210  & 19.4  & 2.72   \\
 2350006886608763008 &  10.559899  & -21.784944  & 18.9  & 2.22   \\
 5001085414388903296 &  10.682683  & -36.794719  & 18.0  & 2.71   \\
 2349905697178722176 &  10.846492  & -22.349365  & 19.4  & 1.83   \\
 2349949578860919040 &  10.867119  & -22.030514  & 19.9  & 2.28   \\
 5001008345495272448 &  10.942463  & -37.263972  & 19.4  & 0.85   \\
 2349122295143983360 &  10.977585  & -22.865027  & 19.0  & 0.95   \\
 2349133427699269888 &  11.004514  & -22.688900  & 18.1  & 2.17   \\
 5000707388546774272 &  11.038949  & -38.288535  & 19.4  & 0.71   \\
 5001186436314116480 &  11.100714  & -36.541186  & 19.5  & 1.91   \\
 5000764631871010432 &  11.106386  & -37.986873  & 19.9  & 1.87   \\
 5000709381411691648 &  11.132341  & -38.212503  & 18.7  & 0.73   \\
 5001112696020754304 &  11.157436  & -36.966320  & 19.3  & 0.84   \\
 5001135102865013248 &  11.253751  & -36.647412  & 19.8  & 0.76   \\
  ... &  ...  & ...  & ...  & ...   \\
\hline
\end{tabular}
\end{center}
\end{table}

\section{The validation of 580V radial velocities} 
\label{sec:580v_rv_validation}

\begin{figure}
    \centering
    \includegraphics[width=0.5\textwidth]{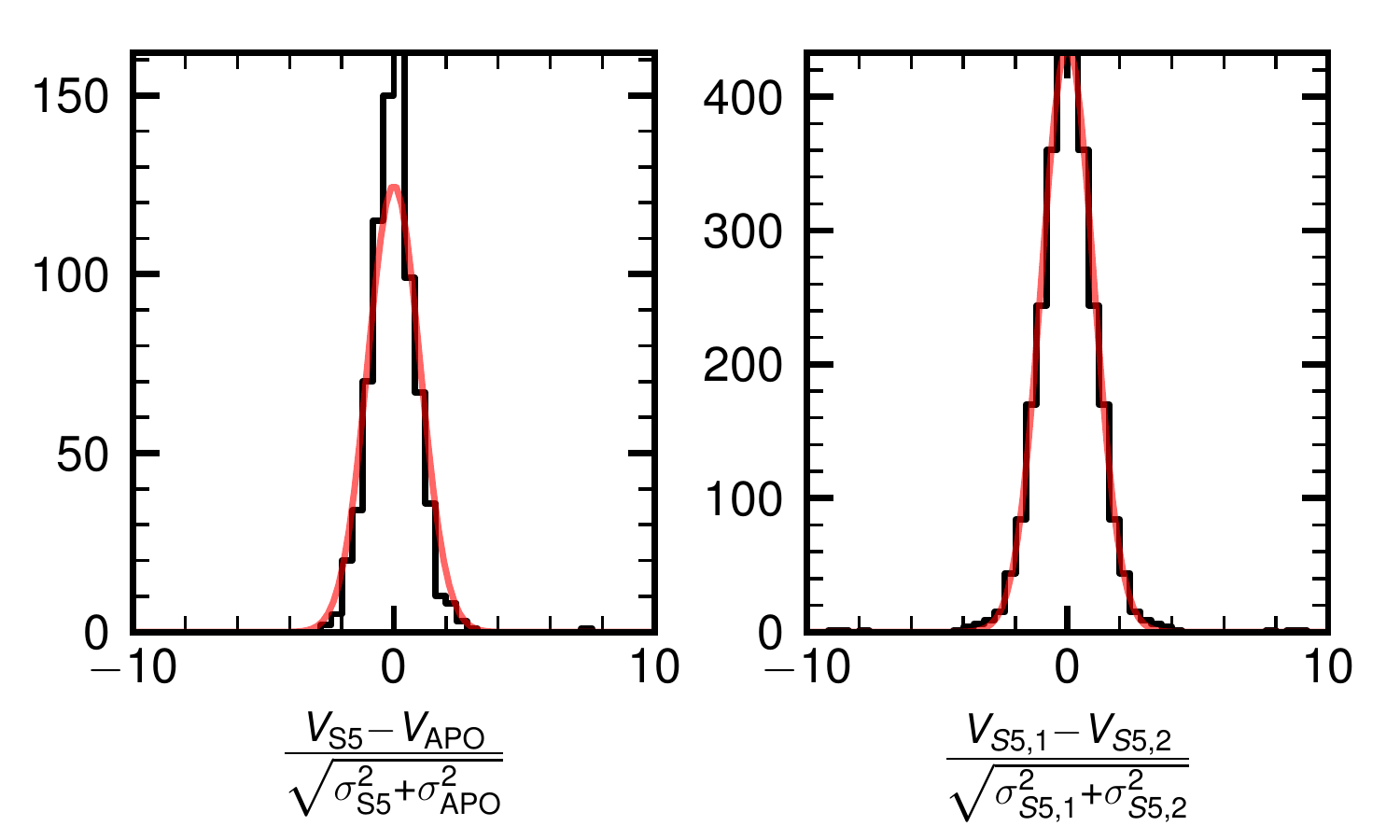}
    \caption{
    Comparison of \SSSSS radial velocities determined from blue arm (580V) spectra with APOGEE (left panel) and repeated \SSSSS observations (right panel). The Figure is identical to Figure \ref{fig:rvcompare} but is using radial velocities determined from the blue arm spectra.
    }
    \label{fig:rvcompare_blue}
\end{figure}

The validation of the measurements from the blue arm (580V grating) requires first identifying spectra that are strongly affected by the bright moon and sky subtraction issues, leading to  the contamination of the spectra by the Solar spectrum (see Section~\ref{sec:reduction}). The issues are particularly prominent in spectra of faint stars. To automatically identify these objects we train a random forest classifier to identify spectra with a large difference between 1700D RVs and 580 RVs, $|v_{580V}-v_{1700D}|>\sqrt{\sigma_{580V}^2+\sigma_{1700D}^2}$ (indicative of 580V sky subtraction issues), using  the moon distance to the field, moon altitude above the horizon and moon phase as features. We train separate classifiers for stars with different \gaia magnitudes: $G<15$, $15<G<16$, $16<G<17$, $17<G<18$, $18<G<19$ and $G>19$. We set the {\tt bad\_moon} flag for stars identified by the classifier.  In total $\sim 10,000$ stars are flagged. 
As expected, none of the bright stars are flagged, while for fainter stars the classifier mostly marks exposures where the moon phase is high and the moon is above the horizon. 

To validate the RVs coming from the 580V spectra we follow the same steps as described in Section~\ref{sec:rv_validation}. We only use stars that are not affected by the Moon (i.e., those do not have the {\tt bad\_moon} flag set). 
This allows us to compute the offset of the velocities as well as calibrate the error model. Due to the much lower spectral resolution of the 580V grating compared to 1700D, as well as the lack of sky-lines for the wavelength calibration of science exposures, the offsets and systematic uncertainties of the 580V RVs are noticeably larger.
We provide them here. The offset is $\sim$ 9\,\kms.
$$
v_\mathrm{S5,580V} = v_\mathrm{rvspecfit}- 8.96 \kms
$$
The RV errors in the blue have a systematic floor of 21.06\kms and the multiplicative factor of $k=1.52$.
Thus our final RV uncertainties in 580V are determined as 
$$\sigma_{v,\mathrm{S5,580V}} = \sqrt{(1.52 \, \sigma_{v,\mathrm{rvspecfit}})^2 + 21.06^2}$$ Figure~\ref{fig:rvcompare_blue} shows the validation of the re-calibrated 580V RVs and uncertainties using APOGEE data and repeated observations (similar to Figure~\ref{fig:rvcompare}).

% Don't change these lines
\bsp	% typesetting comment
\label{lastpage}
\end{document}